\documentclass[letterpaper,prx,aps,twocolumn,showpacs,superscriptaddress,nofootinbib]{revtex4-1}
\usepackage{bm,color}
\usepackage[T1]{fontenc}
\usepackage[utf8]{inputenc}
\usepackage{latexsym}
\usepackage{graphicx} 
\usepackage{amsmath} 
\usepackage{amsfonts} 
\usepackage{braket}
\usepackage{subfigure}
\usepackage{lineno}

\def \be{{\bf e}}

\def \pg{{\phi_{\rm g}}}
\def \er{{E_{\rm rec}}}
\def \bsr{{\boldsymbol{r}}}
\def \bsb{{\boldsymbol{b}}}
\def \bsq{{\boldsymbol{q}}}
\def \bsk{{\boldsymbol{k}}}
\def \bsp{{\boldsymbol{\psi}}}
\def \Vpot{{V_\mathrm{pot}}}

\def \mus{\mu \mathrm{s}}

	\newcommand{\bbm}{\begin{pmatrix}}
	\newcommand{\ebm}{\end{pmatrix}}

\def\be{\begin{equation}}
\def\ee{\end{equation}}

\begin{document}
\title{Multi-frequency optical lattice for dynamic lattice-geometry control}

\author{M. N. Kosch}
\affiliation{Institut f{\"{u}}r Laserphysik, Universit{\"{a}}t Hamburg, 22761 Hamburg, Germany}
\author{L. Asteria}
\affiliation{Institut f{\"{u}}r Laserphysik, Universit{\"{a}}t Hamburg, 22761 Hamburg, Germany}
\affiliation{The Hamburg Centre for Ultrafast Imaging, 22761 Hamburg, Germany}
\author{H. P. Zahn}
\affiliation{Institut f{\"{u}}r Laserphysik, Universit{\"{a}}t Hamburg, 22761 Hamburg, Germany}
\author{K. Sengstock}
\affiliation{Institut f{\"{u}}r Laserphysik, Universit{\"{a}}t Hamburg, 22761 Hamburg, Germany}
\affiliation{The Hamburg Centre for Ultrafast Imaging, 22761 Hamburg, Germany}
\affiliation{Zentrum f{\"{u}}r Optische Quantentechnologien, Universit{\"{a}}t Hamburg, 22761 Hamburg, Germany}
\author{C. Weitenberg}
\email{christof.weitenberg@physnet.uni-hamburg.de}
\affiliation{Institut f{\"{u}}r Laserphysik, Universit{\"{a}}t Hamburg, 22761 Hamburg, Germany}
\affiliation{The Hamburg Centre for Ultrafast Imaging, 22761 Hamburg, Germany}


\begin{abstract}
Ultracold atoms in optical lattices are pristine model systems with a tunability and flexibility that goes beyond solid-state analogies, e.g., dynamical lattice-geometry changes allow tuning a graphene lattice into a boron-nitride lattice. However, a fast modulation of the lattice geometry remains intrinsically difficult. Here we introduce a multi-frequency lattice for fast and flexible lattice-geometry control and demonstrate it for a three-beam lattice, realizing the full dynamical tunability between honeycomb lattice, boron-nitride lattice and triangular lattice. At the same time, the scheme ensures intrinsically high stability of the lattice geometry. We introduce the concept of a geometry phase as the parameter that fully controls the geometry and observe its signature as a staggered flux in a momentum space lattice. Tuning the geometry phase allows to dynamically control the sublattice offset in the boron-nitride lattice. We use a fast sweep of the offset to transfer atoms into higher Bloch bands, and perform a new type of Bloch band spectroscopy by modulating the sublattice offset. Finally, we generalize the geometry phase concept and the multi-frequency lattice to three-dimensional optical lattices and quasi-periodic potentials. This scheme will allow further applications such as novel Floquet and quench protocols to create and probe, e.g., topological properties. 

\end{abstract}

\maketitle

\section{Introduction}

Cold atoms in optical lattices have emerged as a prolific platform to study quantum phases of, e.g., solid-state models \cite{Lewenstein2012} with their great advantage of dynamic tunability. Of particular current interest are non-separable and bipartite lattices \cite{Windpassinger2013} such as superlattices \cite{Anderlini2007,Trotzky2008}, checkerboard lattices \cite{Wirth2011}, triangular lattices \cite{Becker2010, Struck2011}, honeycomb lattices \cite{Becker2010,Tarruell2012,Flaschner2016}, Lieb lattices \cite{Taie2015}, Kagome lattices \cite{Jo2012} or quasicrystal lattices \cite{Viebahn2019}. As the lattice potential is created artificially with interfering light beams, dynamical changes between those different lattice types or specific lattice geometries are in principle possible and allow for fascinating scenarios, far beyond condensed matter options, e.g., adiabatic or fast tuning from one lattice geometry to another. However a tuning possibility technologically typically contrasts with the lattice stability, necessary to avoid heating of the atoms.

In general, in $d$ dimensions, $d+1$ laser beams result in an intrinsically stable lattice geometry \cite{Petsas1994}, which is, however, also intrinsically static, i.e., non-tunable. More precisely the lattice-beam polarization determines the lattice geometry \cite{Sebby-Strabley2006,Becker2010,Baur2014,Flaschner2016} but can typically only be changed slowly. More laser beams or frequencies allow for a tunable lattice geometry \cite{Anderlini2007,Trotzky2008,Tarruell2012,Dai2016,Robens2017,Wang2021,Li2021} but so far require phase locks for control and stability. 

\begin{figure}[!]
	\includegraphics[width=0.90\linewidth]{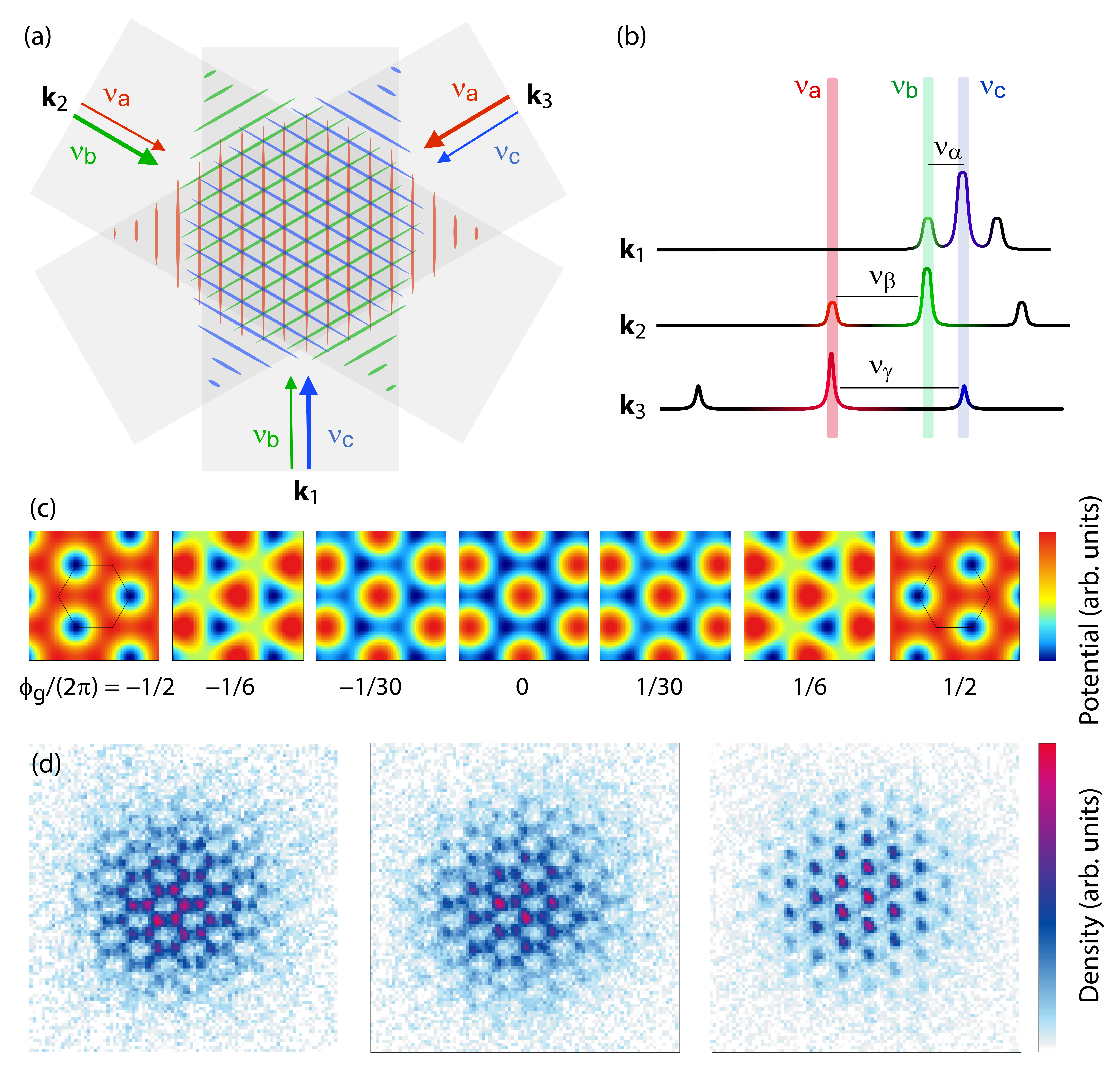}
	\caption{Realization of a hexagonal lattice with tunable and stable geometry as multi-frequency lattice. (a) In the multi-frequency design a hexagonal lattice is realized by superimposing three 1D lattices of different laser frequencies  indicated as red ($\nu_a$), green ($\nu_b$) and blue ($\nu_c$). Bold arrows represent carriers, thin arrows represent sidebands of the three lattice beams with wavevectors $\boldsymbol{k}_1$, $\boldsymbol{k}_2$, and $\boldsymbol{k}_3$. (b) The beam configuration in (a) can be obtained by appropriate sidebands ($\nu_\alpha$, $\nu_\beta$, $\nu_\gamma$) imprinted on the laser beams together with suitable detunings between the beams. (c) The resulting lattice can be freely tuned between a honeycomb lattice (center), boron-nitride lattices, and triangular lattice (edges). The numbers below the plots indicate the corresponding geometry phase, the hexagons show a unit cell. (d) Clouds of 30,000 $^{87}$Rb atoms in different lattice potentials ($\pg/(2\pi)=0,~0.005,~0.03$, from left to right).}
    \label{fig:scheme}
\end{figure}

Here, we introduce the concept of the multi-frequency lattice, which allows to combine both, stability and tunability, and which can be realized in different configurations in two or three dimensions. 
We introduce the geometry phase for description of optical lattices and quasicrystal lattices in any dimension and show how this quantity can be controlled via the multi-frequency scheme. This makes a vital contribution to the general understanding of the tunability of optical lattices. Next to the fast and stable control of hexagonal lattices, which we demonstrate experimentally, we also discuss how to realize similar control in 3D lattices and quasicrystal lattices for the first time. These concepts will be crucial for fully exploiting the potential of optical lattices including quench and Floquet protocols in non-standard lattices.

\section{Geometry phase in three-beam lattices}

We start our explanation of the multi-frequency optical lattice by a general consideration which applies to three-beam lattices. This leads us to introduce as a new concept a geometry phase that fully determines the lattice geometry as explained below. This concept is then extended to general 3D lattices in Appendix A and to quasicrystal lattices in Appendix B.

Each 2D optical lattice formed by three interfering lattice beams with wavevectors $\boldsymbol{k}_i$  ($i=1,2,3$) can be written, without loss of generality, as a sum of three one-dimensional (1D) lattices (see Appendix C):
\begin{equation}
\Vpot(\bsr)=V_0 + 2\sum_{i=1}^3 V_i\cos\Bigl(\bsb_i\cdot\bsr+\phi_i\Bigr) \label{eq:3beam}
\end{equation}
where $V_0$ is an energy offset, $\bsb_i$ are the reciprocal lattice vectors (defined as $\bsb_i=\boldsymbol{k}_i-\boldsymbol{k}_{i+1}$, with $\boldsymbol{k}_4=\boldsymbol{k}_1$), and $\phi_i$ are referred to as the phases of the three 1D lattices, which are determined by the relative phases and polarization of the lattice beams. $V_i$ is the lattice depth of the corresponding 1D lattice and in the symmetric case, $V_1=V_2=V_3=V$ we denote $V$ as the lattice depth. It can be shown that for a fixed triple $V_i$, the three-dimensional space spanned by the $\phi_i$ can be decomposed into two degrees of freedom associated with the lattice position in the 2D plane and one degree of freedom which determines the geometry of the lattice, i.e., the choice of a honeycomb, boron-nitride or triangular lattice in the case of a hexagonal lattice. We demonstrate (see Appendix C) that the relevant parameter is given by: 
\begin{equation}
    \phi_g=\sum_{i=1}^3\phi_i \label{eq:pg}
\end{equation}  
which we refer to as the geometry phase. $\pg$ cannot be changed by phase shifts applied on the laser beams, as these only change the lattice position. A geometry phase can be defined for higher numbers of 1D lattices forming the potential, and it can be in general a higher-dimensional object, e.g. the geometry phase for a three-dimensional (3D) lattice formed by interference of four beams is described by a three-vector (Appendix A). 

\section{The multi-frequency lattice}

Based on the insight that a full control of the geometry phase $\pg$ demands for a full control over the individual phases $\phi_i$ of the 1D lattices, we develop a new, tunable lattice scheme. It is realized by establishing pairwise interference at different frequencies via sidebands modulated onto appropriately detuned laser beams (Fig.~\ref{fig:scheme}). The lattice so formed inherits the geometry stability of 2D lattices made up by three lattice beams \cite{Petsas1994} and allows dynamic control of the lattice geometry via the relative phase of the radio frequency (RF) sidebands, which can be controlled with high precision (see Appendix E).

Our implementation of this multi-frequency lattice features a tunability on the microsecond scale and a high passive stability of the geometry phase with only very slow drifts characterized by a standard deviation of $\delta\phi_\mathrm{rf}^0=0.17^{\circ}$ (see Appendices F, G). Our new lattice scheme allows us to tune the lattice geometry between a honeycomb-(graphene)-lattice, a boron-nitride lattice and a triangular lattice as demonstrated by single-site-resolved images of $^{87}$Rb atoms in the lattice via the recently introduced quantum gas magnifier \cite{Asteria2021} [Fig~\ref{fig:scheme}(d)].

\begin{figure}[!]
	\includegraphics[width=0.8\linewidth]{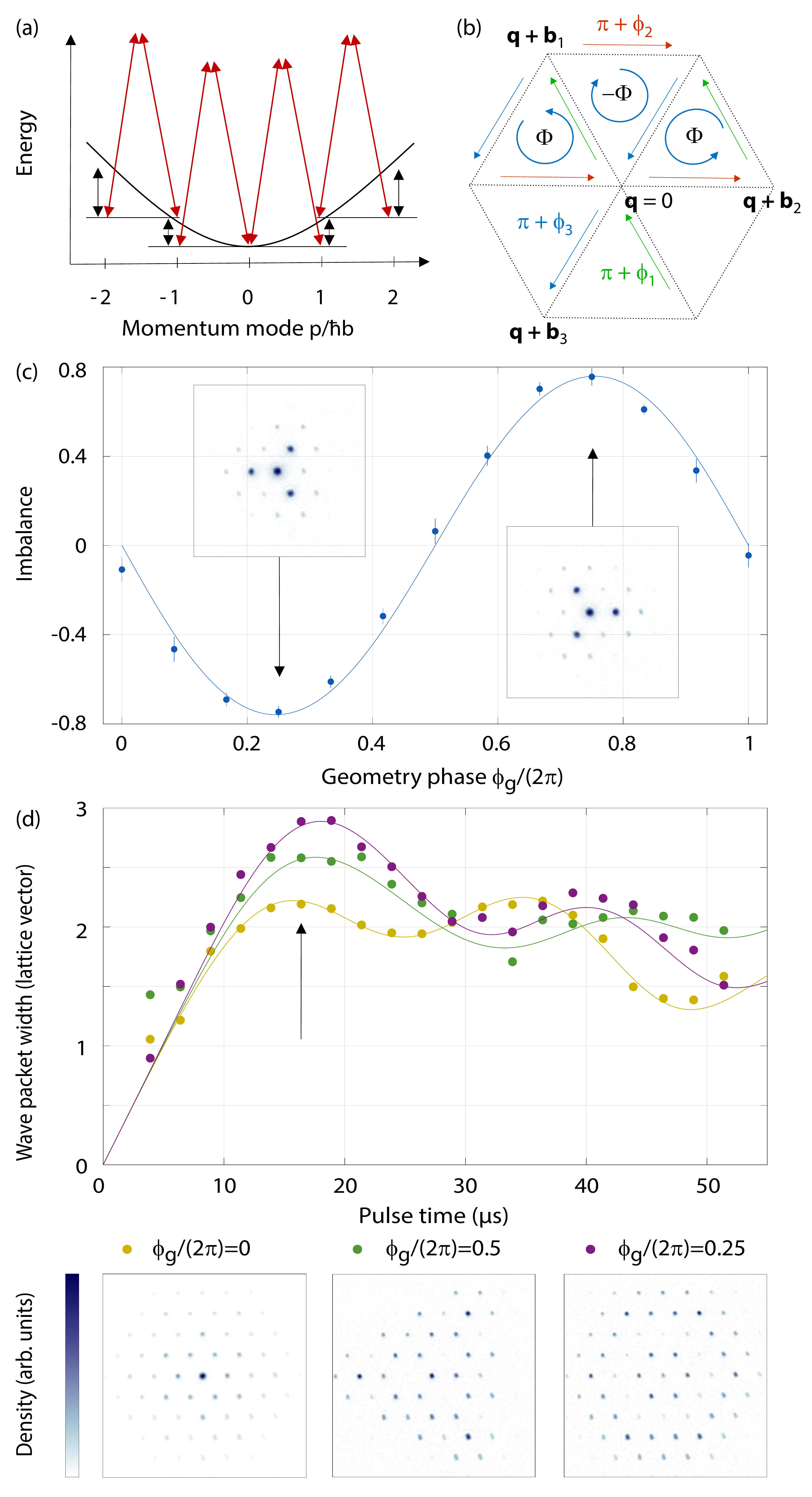}
	\caption{Quantum walk in a triangular momentum space lattice with staggered flux showing inversion-symmetry breaking.
(a) A 1D optical lattice couples momentum modes separated by its wave vector $\boldsymbol{b}$ (red arrows). The coupling becomes increasingly off-resonant for higher momentum modes (black arrows). (b) The resulting lattice in momentum space has a staggered flux $\Phi$ resulting from the Peierls phases for the directions indicated by the arrows. The Peierls phases and the corresponding arrows are displayed in the same color. (c) Imbalance $\mathcal{I}=\frac{p_a-p_o}{p_a+p_o}$ between the 1st order peak populations along ($p_a$) and opposite ($p_o$) to the reciprocal lattice vectors $\bsb_i$ as a function of the geometry phase. The error bars denote the standard deviation of the mean of the three equivalent pairs. The insets show the raw images for the points indicated by the arrows. The images were taken for an effective pulse time of 4.7~$\mus$. (d) Top panel, time evolution of the width of the distribution of experimental data (symbols) and numerics (lines) for $\pg/2\pi=0$ (yellow), $\pg/2\pi=0.25$ (green) and $\pg/2\pi=0.5$ (purple). The external trap leads to halting of the expansion, with different exact behaviour as a function of the staggered flux. Exemplary distributions at a pulse time of 16.4~$\mus$ (indicated by the arrow) show the rich dynamics in dependence of the staggered flux (lower panels). The lines in (c) and (d) are numerically calculated using $V=6.5 E_{\rm rec}$ (Appendix H).}
    \label{fig:quantum-walk}
\end{figure}

\section{Staggered flux in momentum-space lattice}

To further demonstrate the relevance of the geometry phase concept, we experimentally study its effect in a momentum-space lattice \cite{Gadway2015,Meier2016}, using our ability to tune it via the multi-frequency lattice. The geometry phase appears in the triangular momentum-space lattice formed by all momenta differing by reciprocal lattice vectors. The Bragg scattering of two lattice beams imprints their relative phase, i.e., the phase $\phi_i$ of the respective 1D lattice in Eq.~(\ref{eq:3beam}). This yields Peierls phases $\phi_i+\pi$ [Fig.~\ref{fig:quantum-walk}(b)], where the shift by $\pi$ stems from the convention of a minus sign in tight-binding lattice Hamiltonians. We consider the staggered flux through the triangular plaquettes $\Phi$, i.e., the sum of the Peierls phases along a plaquette, which is given by
\begin{equation}\Phi = \sum_i^3\;(\pi+\phi_i) = \pi + \pg \mathrm{.}
\end{equation} 
This shows that while the Peierls phases depend on the choice of the origin, the staggered flux is independent of this choice, because it is only determined by the geometry phase.

We realize a continuous-time quantum walk in the triangular momentum space lattice \cite{Gadway2015} in order to probe this staggered flux [Fig.~\ref{fig:quantum-walk}(a),(b)]. The quantum walk is realized by pulsing on the lattice as Kapitza-Dirac scattering and measuring the occupation of the different momentum modes (Bragg peaks) after time-of-flight expansion (Appendix H). We observe the presence of a flux manifested in the breaking of inversion symmetry for values of $\pg$ different from $\pg=0$ (honeycomb lattice) or $\pg=\pi$ (triangular lattice) [Fig.~\ref{fig:quantum-walk}(c)]. The symmetry breaking can be understood to arise from the alternating pattern of currents, which favor movement into three of the six neighbouring modes [Fig.~\ref{fig:quantum-walk}(b)]. This interpretation of a staggered flux in momentum space also provides a physical picture for earlier measurements of Kapitza-Dirac scattering in honeycomb lattices \cite{Thomas2016, Weinberg2016PRA}. While the real-space densities are sensitive to the geometry phase around $\pg=0$ [Fig.~\ref{fig:scheme}(d)], these momentum-space measurements allow us to sensitively probe the geometry phase in its entire range. A similar observation was made in a triangular lattice from circular polarized beams, which is effectively triangular in real space, but shows pronounced symmetry breaking in Kapitza-Dirac scattering \cite{Yang2021}.

The tunable staggered flux appears naturally in the momentum-space lattice, while it was actively created via Floquet engineering in a real-space triangular lattice \cite{Struck2011}. While such real-space lattices allow exploring interacting systems in the ground state, which can additionally lead to spontaneous symmetry breaking at $\Phi=\pi$ \cite{Struck2013}, momentum-space lattices are ideally suited to study dynamics, e.g., starting from a single site.

Our setup differs from previous realizations of a momentum space lattice as artificial dimension \cite{Gadway2015,Meier2016}, where the transitions between the momentum modes are coupled via individual frequency components of the laser beams allowing to compensate for the increasing kinetic energy of the higher momentum modes. In our case of global frequencies, the increasing energy mismatch leads to an effective harmonic trap in momentum space [Fig.~\ref{fig:quantum-walk}(a)]. Setups with individual frequency components also allow creating rectified magnetic fluxes as demonstrated for momentum-space flux ladders \cite{An2017} and proposed for triangular spin-momentum lattices \cite{Lauria2022}. In contrast, our experiment allows making a connection to the Bloch coefficients of the real-space optical lattice, which dictate the occupations of the momentum modes of the ground state. The difference of the Bloch coefficients of a honeycomb and triangular lattice, can then be thought of as arising from the staggered flux $\Phi$. 

The effect of the harmonic trap can be clearly observed in the dynamics: the initial expansion halts and the width of the distribution oscillates around a plateau of about 2 momentum modes while forming interesting patterns that depend on the value of the staggered flux [Fig.~\ref{fig:quantum-walk}(d)]. While a rectified flux would also lead to a spatial confinement of a quantum walk \cite{Chalabi2019}, in our case the halting of the expansion is an effect of the trap.

\section{Preparation of higher bands via sweeps of the geometry phase}

\begin{figure}
	\includegraphics[width=0.95\linewidth]{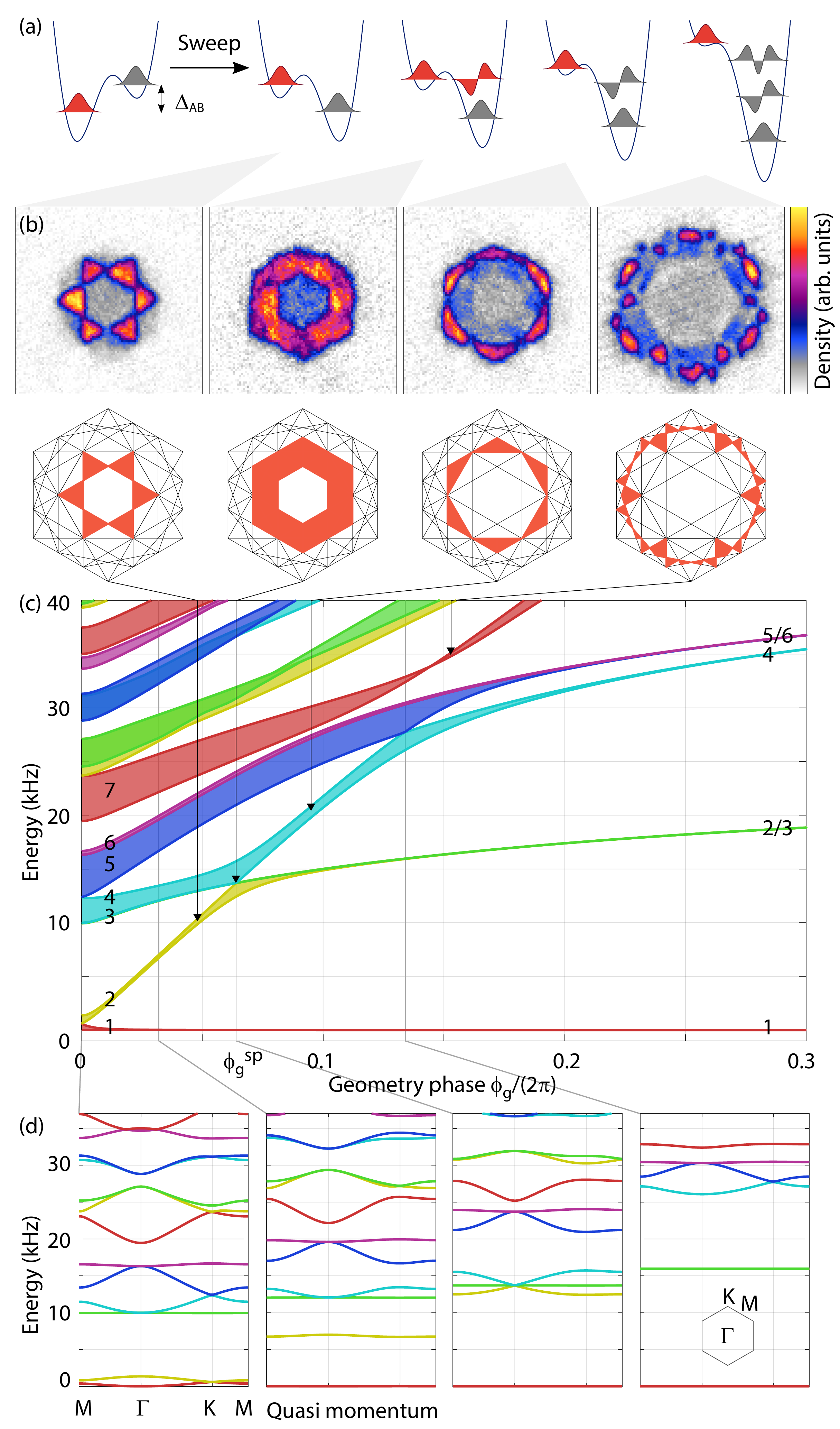}
	\caption{Preparation of higher bands via sweeps of the geometry phase. (a) 1D cuts through the potentials before and after the sweeps inverting the energy offset between the A and B sites $\Delta_{\rm AB}$. The populated orbitals are highlighted in red. The four final situations correspond to the four panels in (b). (b) Band mapping images for linear sweeps from $\pg/(2\pi)=-0.048$ to final values of $\pg/(2\pi)=0.048$, $0.064$, $0.095$, and $0.153$ within $60~\rm{\mu s}$ ($20~\rm{\mu s}$ in the last image) followed by $500~\rm{\mu s}$ hold time. The atoms are transferred to the 2nd band, 2nd+3rd+4th band (at the band touching), 4th band, and 7th band, respectively. Below the Brioullin zones corresponding to these bands are shown (c) Band structure of the hexagonal lattice as a function of $\pg$. The band structure is symmetric under reflection around $\pg=0$ and $\pg=\pi$ (not shown). The lowest band gap given by $\Delta_\mathrm{AB}$ increases linearly with $\pg$ up to about $\pg^\mathrm{sp}$, where the energetically highest s-orbital is degenerate with the lowest p-orbital ($\pg^\mathrm{sp}/(2\pi)=0.064$ for our lattice depth of $V=6.3~\er$). Via a series of (avoided) crossings, the bands rearrange from a honeycomb lattice with two s-bands ($\pg=0$) to a triangular lattice ($\pg=\pi$) with a single s-band. Arrows indicate the final values of $\pg$ of the sweeps stated in (b). (d) Plots of the band structures at $\pg/(2\pi)=0$, $0.032$, $0.064$, and $0.134$, the latter two featuring triple band crossings.}
    \label{fig:bandstructure}
\end{figure}

The geometry phase provides a unified description of the transition from honeycomb to triangular lattice with a single tuning parameter [Fig.~\ref{fig:bandstructure}(c)]. The different order of even and odd bands between the Bravais lattice and the lattice with two-atomic basis enforces a series of band crossings as a function of $\pg$. These crossings produce interesting band structures with Dirac points and triple crossings between higher bands [Fig.~\ref{fig:bandstructure}(d)]. At specific values of $\pg$, the band structures contain interesting triple band crossings reminiscent of the Lieb lattice between the 2nd to 4th band and reminiscent of the Kagome lattice between the 4th to 6th band. Such relatively simple realizations of these band structures could be employed to probe the corresponding geometric properties via wavepacket dynamics \cite{Brown2022}. 

The full dynamic control of the lattice allows the transfer into higher bands as previously used for checkerboard lattices \cite{Wirth2011,Olschlager2011,Kock2015,Hachmann2021} and hexagonal lattices \cite{Weinberg2016, Jin2021, Wang2021}. Here we prepare the atoms in higher bands by appropriate sweeps of $\pg$ and detect the result via band mapping [Fig.~\ref{fig:bandstructure}(a,b)]. The precise control of higher bands in hexagonal lattices is important for accessing exotic higher-band physics \cite{Wu2008}, such as the chiral superfluid recently realized \cite{Wang2021}. In the future, it would be interesting to directly image the higher orbitals and the on-site vortices of the chiral superfluid using a quantum gas magnifier \cite{Asteria2021}.

\section{Sublattice modulation spectroscopy} 

Using the dynamical control of the geometry of the lattice, we have access to an additional method for modulation of the lattice potential. When modulating the geometry phase, we realize a modulation of the on-site energy difference between the sublattices, which we refer to as sublattice modulation spectroscopy. 

\begin{figure}
	\includegraphics[width=0.95\linewidth]{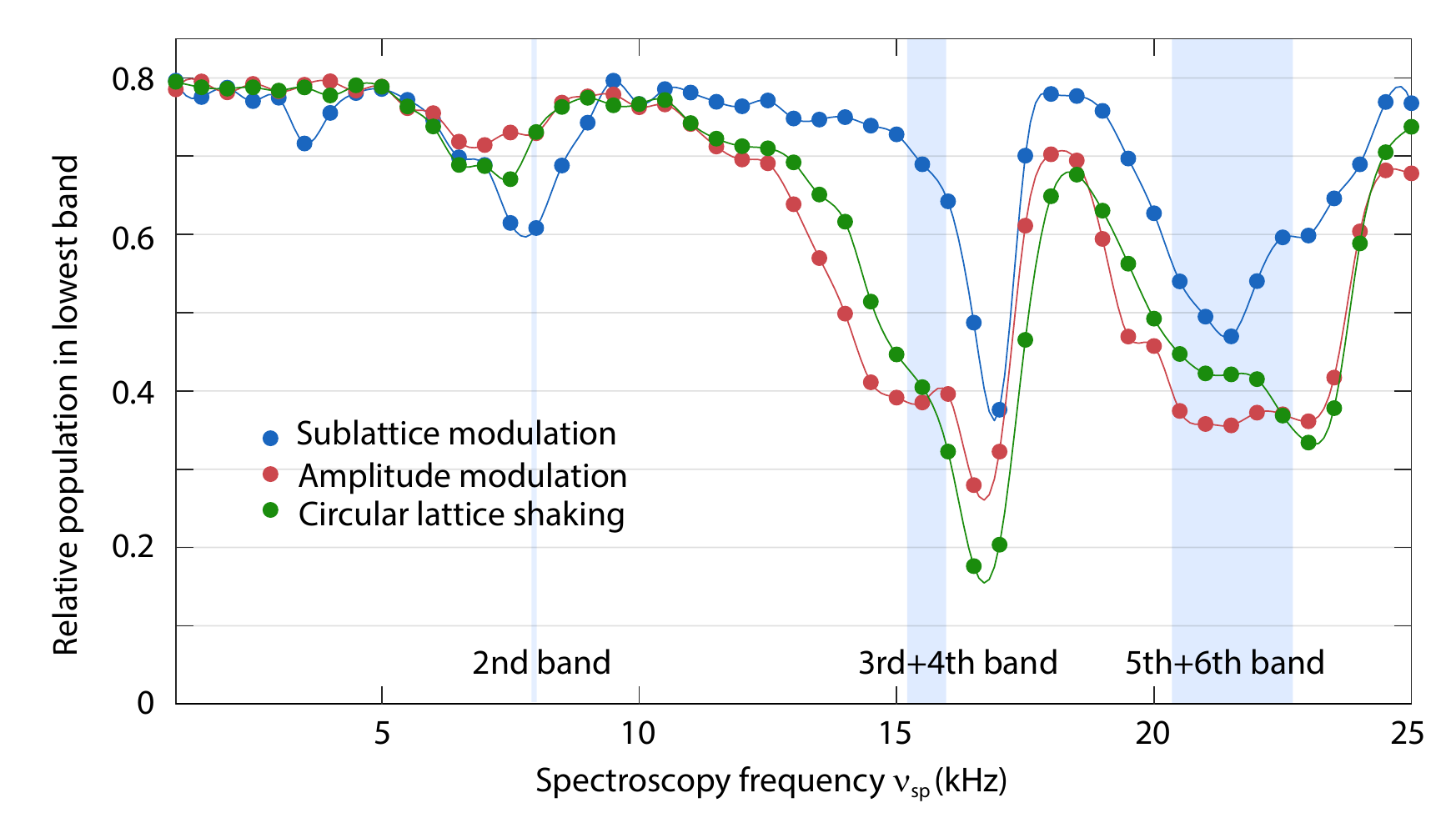}
	\caption{Excitation spectra of a boron-nitride lattice with $\pg/(2\pi) = 0.024$ for different modulation types. Relative population of the lowest band obtained via band mapping versus spectroscopy frequency $\nu_\mathrm{sp}$ for sublattice modulation (blue), amplitude modulation (red) and circular lattice shaking (green). The lines are spline interpolations as a guide to the eye. The modulation indices are 0.02, 0.05 and $600$\,Hz$/\nu_{\rm sp}$ respectively, chosen to yield good signal at the modulation time of 10\,ms. The shaded areas indicate the expected transition frequencies into the stated bands for a lattice depth of $V_{\rm latt}=9.3\,E_{\rm rec}$.} 
    \label{fig:spectroscopy}
\end{figure}

We exemplify this sublattice modulation by multi-band spectroscopy of $^{87}$Rb atoms in a boron-nitride lattice and compare it to amplitude modulation \cite{Flaschner2018} and circular lattice shaking \cite{Asteria2019, Weinberg2015} (Fig.~\ref{fig:spectroscopy} and Appendices I, J, K). We find that sublattice modulation yields cleaner spectra than the other two modulation types in the sense that it features a broad region without excitations, which could be used for off-resonant Floquet protocols \cite{Eckardt2017,Weitenberg2021}. 

Sublattice modulation spectroscopy was so far not realized in optical lattices and extends the possibilities of lattice modulation. Related phasonic modulation via an incommensurate secondary moving lattice was previously found to efficiently couple via high-order multi-photon transitions \cite{Rajagopal2019}. Similar superlattice modulation spectroscopy was predicted to be advantageous for probing the bond order wave in the ionic Hubbard model by accessing both its finite charge and spin gaps \cite{Loida2017}. Further analysis could be done to identify many-body systems, where sublattice modulation selectively couples to the quasi-particle excitations.

\section{Outlook}

In conclusion, we have proposed and implemented a multi-frequency optical lattice, which combines full dynamic tunability and high stability of the lattice geometry. We expect this multi-frequency approach to be very useful for a large range of experiments beyond the applications demonstrated here. Fast dynamic control is crucial for quenches onto different lattice geometries, as employed in Bloch state tomography for characterizing topological properties \cite{Hauke2014,Flaschner2016, Tarnowski2019} or for detection protocols of many-body phases \cite{Nuske2016, Zheng2020}. The state tomography could now be realized for a wide variety of systems, because the sublattice offset after the quench can be chosen independently of the system under consideration. Dynamically changing the potential from triangular to honeycomb lattice could be used to add a second potential well on each site, making possible an in-situ Stern Gerlach separation for spin-resolved read out \cite{Boll2016} also in a triangular lattice. 

The direct modulation of the geometry allows novel spectroscopy methods as well as Floquet protocols, which make use of reduced heating rates from reduced coupling to higher bands or which utilize the inversion-symmetry breaking induced by sublattice-modulation to study the influence on topological phases. Furthermore, the method gives the freedom to tune the geometry independently of the lattice beam polarization allowing to completely avoid vector light shifts or to employ them at will.
The multi-frequency design could be used to extend the fast and tunable control demonstrated here for 2D lattices also to 3D lattices (Appendix A) or to quasicrystal lattices (Appendix B), which would enable Floquet engineering of new topological phases \cite{Huang2021_hingestates}.
Finally, the scheme might be used to create a spatially varying lattice geometry, e.g., for engineering topological interfaces (Appendix D).

\section*{Acknowledgments} 

The work was funded by the Cluster of Excellence 'CUI: Advanced Imaging of Matter' of the Deutsche Forschungsgemeinschaft (DFG) - EXC 2056 - project ID 390715994, by the DFG Research Unit FOR 2414, project ID 277974659 and by the European Research Council (ERC) under the European Union's Horizon 2020 research and innovation programme under grant agreement No. 802701. 


\section*{Appendix A: Generalization to 3D lattices}

\begin{figure}[!]
\includegraphics[width=0.80\linewidth]{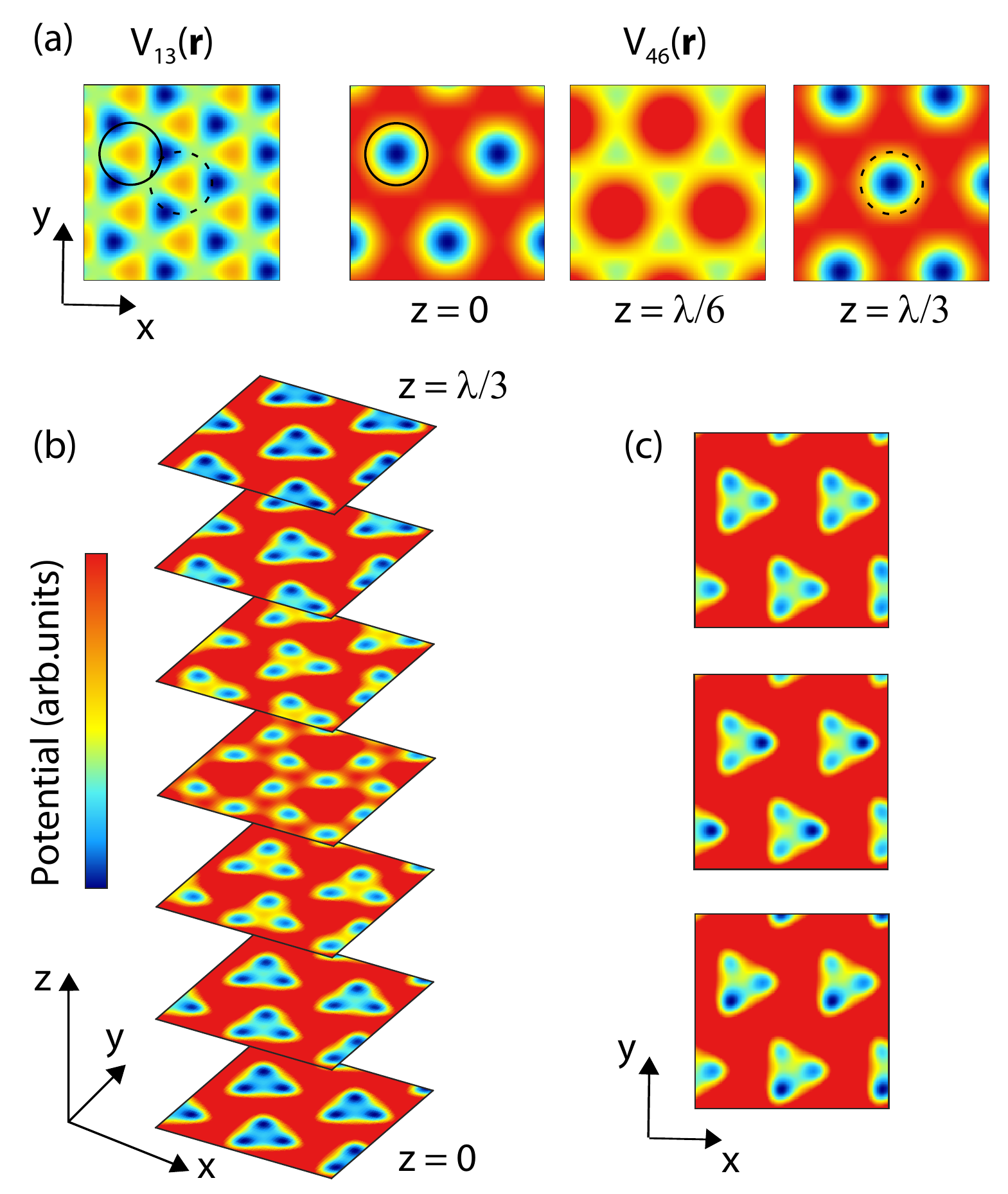}
\caption{Proposal for a tunable 3D lattice of trimers via multi-frequency design. 
(a) $V_{13}(\bsr)$ and $V_{46}(\bsr)$ for $z=0,\,\lambda/6,\,\lambda/3$, as defined in the main text. The geometry phase ($\phi_{g,1}/(2\pi)=0.29$, $\phi_{g,2}=\phi_{g,3}=\phi_{g,1}/3$) is chosen such that the minima of $V_{46}(\bsr)$, found at $z=m\lambda/3$ (with $m$ integer), coincide with a maximum of $V_{13}(\bsr)$ (solid and dashed circle). The relative depth of the two potentials is given by $V_4=1.8V_1$.
(b) Potential landscape plotted for different sections parallel to the $xy$ plane of size $2.2\lambda\times2.2\lambda$. The colormap is cropped to emphasize the relevant potential minima. The lattice has three sublattices, i.e., trimers in the $xy$ plane, and tunneling along $z$ couples lattice sites in different sublattices.
(c) Potential in the plane of the trimers with different potential offsets between the sublattices, engineered by shifting $V_{46}(\bsr)$ with respect to $V_{13}(\bsr)$ in the $xy$ plane using different choices of the geometry phases $\phi_{g,2}$ and $\phi_{g,3}$ [compare Fig.~\ref{fig:trimers}(a)]. A balanced situation is achieved without shift (first panel) and sublattice offsets are reached by shifting along the $x$ direction (second panel) and the $y$ direction (third panel).}
    \label{fig:trimers}
\end{figure}

3D non-separable optical lattices for ultracold atoms have not been realized so far, yet 3D systems feature Weyl semimetals \cite{Yan2017,Wang2021Science}, higher order topological insulators \cite{Benalcazar2017,Schindler2018b} and can be used to simulate a wider class of solid-state materials. We show here how the multi-frequency design can be extended to 3D lattices providing a high degree of control over the lattice geometry without the technical difficulties and constraints of the polarization design approach. Note that in 3D the polarization choice is non-trivial and need, e.g., polarization optimization algorithms \cite{Cai2002}. Additionally, a 3D multi-frequency lattice would also allow for a dynamic variation of the geometry in Floquet protocols \cite{Huang2021_hingestates}.

A 3D lattice can be realized by interference of four non-coplanar beams with wavevectors $\boldsymbol{k}_i$, with $i=1,2,3,4$. The resulting lattice potential will have spatial Fourier components at 6 wavevectors $\boldsymbol{b}_{i}$, defined as before for $i=1,2,3$ and defined as  $\boldsymbol{b}_i=\boldsymbol{k}_4-\boldsymbol{k}_{i-3}$ for $i=4,5,6$. The total potential can be written as:
\begin{equation}
\Vpot(\bsr)=V_0+2\sum_{i=1}^6V_{i}\cos(\boldsymbol{b}_{i}\cdot \boldsymbol{r}+\phi_{i})    
\end{equation}
The potential is characterized in this case by six degrees of freedom associated with the phases $\phi_i$. Three of them are coupled to the position of the lattice in space, and hence the remaining three are needed for describing the lattice geometry. A possible choice of an independent triple is e.g.:
\begin{equation}
    \begin{aligned}
    \boldsymbol{\phi}_g=
 \begin{bmatrix}
\phi_{g,1} \\
\phi_{g,2} \\
\phi_{g,3}
\end{bmatrix}=
 \begin{bmatrix}
\phi_1+\phi_2+\phi_3 \\
\phi_1+\phi_4-\phi_5 \\
\phi_2+\phi_5-\phi_6 
\end{bmatrix}
  \end{aligned}\label{eq:pg3D}
\end{equation}
where each component is constructed such that it is independent of the phases of the lattice beams; ${\phi}_{g,1}$ is the geometry phase for the 2D lattice obtained with beams of wavevectors  $\bsk_1,~\bsk_2,~\bsk_3$, ${\phi}_{g,2}$ is the geometry phase for the 2D lattice obtained with beams of wavevectors $\bsk_1,~\bsk_4,~\bsk_2$, and ${\phi}_{g,3}$ is the geometry phase for the 2D lattice obtained with beams of wavevectors $\bsk_2,~\bsk_4,~\bsk_3$. Equivalently, also $\phi_3-\phi_4+\phi_6$ is the geometry phase of a 2D lattice, but it is not independent from the others as it could be obtained as $\phi_{g,1}-\phi_{g,2}-\phi_{g,3}$. 

The implementation of a 3D multi-frequency lattice can be accomplished via suitable sidebands on the laser beams such that each beam pair creating a 1D lattice shares a common frequency, not present in the other beams, in analogy to the 2D lattice demonstrated in the main text. All components of $\pg$ could be calibrated then for each of the three corresponding 2D lattices using the methods demonstrated in this article. Note that in order to control all six $V_i$ depths, not only the four laser intensities would have to be changed but also the strength of the modulation producing the frequency sidebands. A suitable basis transformation within this 3D parameter space might be appropriate to develop an understanding of the tunable geometry of the particular 3D lattice under study.

An overview of all possible geometries obtainable is left for future work, because the parameter space is huge: with an appropriate choice of beam wavevectors, all 14 3D Bravais lattices can be realized \cite{Cai2002}, and also multi-atomic bases such as the diamond lattice \cite{Petsas1994,Toader2004}. As an example, we present here a geometry characterized by three local minima in the unit cell which could be realized in our setup by, e.g., adding a fourth beam pointing in $z$ direction, with wavevector $\boldsymbol{k}_4=\frac{2\pi}{\lambda}\hat{z}$, with vector $\hat{z}$ perpendicular to the $xy$ plane (Fig.~\ref{fig:trimers}).  An interference of all beams could be realized, e.g., via in-plane linear polarization of the first three beams and circular polarization of the fourth beam.
It is convenient, for visualization, to write the total potential as a sum of two terms: $\Vpot(\bsr)=V_{13}(\bsr)+V_{46}(\bsr)$, where
\begin{equation}\begin{aligned}
V_{13}(\bsr)&=2V_1\sum_{i=1}^3\cos(\bsb_{i}\cdot \bsr+\phi_{i})\\ V_{46}(\bsr)&=2V_4\sum_{i=4}^6\cos(\bsb_{i}\cdot \bsr+\phi_{i}),
\end{aligned}
\end{equation}
assuming $V_1=V_2=V_3$ and $V_4=V_5=V_6$, and leaving out constant terms.

From $V_{46}(\bsr)=2\sum_{i=4}^6V_{i}\cos(\frac{2\pi}{\lambda}z-\bsk_{i-3}\cdot \bsr+\phi_{i})$, i.e., the argument of all the cosines in $V_{46}(\bsr)$ has a linear dependence on $z$, it can be seen that horizontal cuts of $V_{46}(\bsr)$ look like a hexagonal lattice in the $xy$ plane with a geometry phase $\pg^z$ given by $\pg^z=3z\frac{2\pi}{\lambda}$. The horizontal cuts of $V_{46}(\bsr)$ possess a three times bigger unit cell than $V_{13}(\bsr)$, rotated by 30°.

The lattice vectors of the 3D lattice $\boldsymbol{a}_1,\;\boldsymbol{a}_2,\;\boldsymbol{a}_3$ can be identified from the distance between the minima of $V_{46}(\bsr)$ as:
\begin{equation}
    \boldsymbol{a}_i=\frac{2\lambda}{3}\sin(\frac{2\pi}{3}i)\hat{x}+\frac{2\lambda}{3}\cos(\frac{2\pi}{3}i)\hat{y}+\frac{\lambda}{3}\hat{z}
\end{equation}
where $\hat{x},\,\hat{y},\,\hat{z}$ are the unit vectors of the corresponding directions. The resulting potential can be described as layered planes with tunnel-coupled trimers, where each lattice site is only coupled to a site of the neighboring trimer along the perpendicular direction.
Because the 3D lattice is non-separable, the trimers are also not trivially coupled between adjacent planes and tunneling always couples different sublattices.

Furthermore, the lattice parameters are tunable and, e.g., offsets between the sublattice sites can be engineered by control of $\phi_{g,2}$ and $\phi_{g,3}$ [Fig.~\ref{fig:trimers}(c)].
Balanced ratios of tunnel elements along the different directions can be achieved by tuning the $V_i$ or via an out of plane component of the three first beams, similar to tetrahedral laser beam configurations previously explored with thermal atoms \cite{Verkerk1994,Birkl1995}.

\begin{figure*}
	\includegraphics[width=0.85\linewidth]{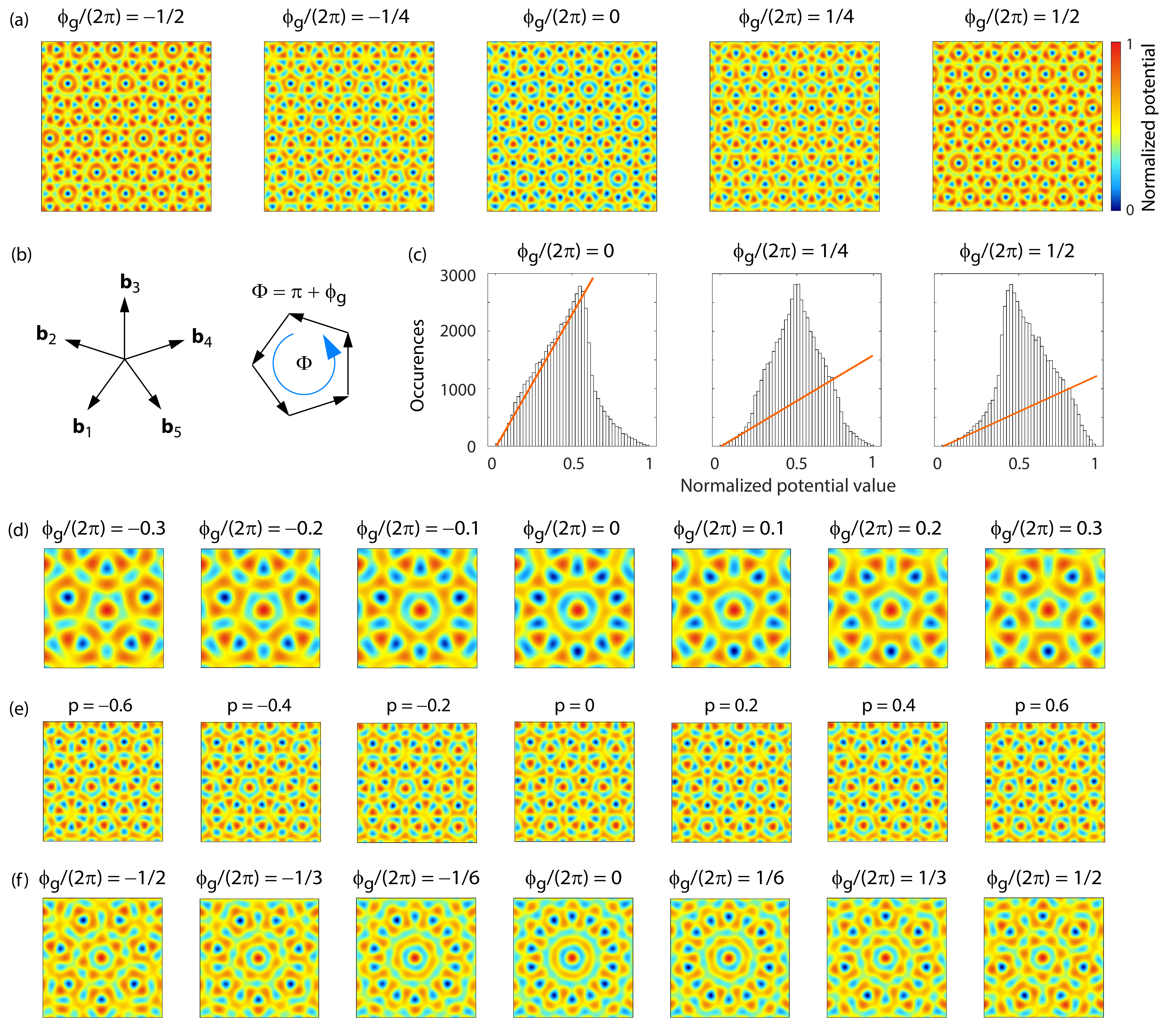}
	\caption{Geometry phase in a quasicrystal lattice.  
	(a) Lattice potential of the form (Eq. \ref{eq:potential}) as a function of $\pg$ in an area of $10\lambda\times10\lambda$. The lattice potential is normalized to the interval [0,1]. (b) Orientation of the reciprocal lattice wavevectors $\boldsymbol{b}_i$, with $|\boldsymbol{b}_i|\sim1.9\frac{2\pi}{\lambda}$. In momentum space, pentagonal plaquettes with flux $\Phi=\pm(\pg+\pi)$ can be found.  (c) Histograms of the values of the normalized potential landscape  for $\pg/(2\pi) = 0,\,1/4,\,1/2$. Orange lines are drawn to mark the slope at the smallest potential values, i.e. in the relevant region for the lowest lying states. The slope is largest at $\pg=0$ because of the larger number of lattice sites of similar energy.
	(d) Zoom in ($2.58\lambda\times2.58\lambda$) on one of the 10-fold symmetric pattern that can be found for $\pg=0$, demonstrating the breaking of the symmetry between the odd and the even minima of the structure for finite values of $\pg$. 
	(e) Normalized quasicrystal potential in a $5\lambda\times5\lambda$ area as a function of a phasonic degree of freedom $p$: $\Vpot(\bsr)=\sum_i^5 \cos(\bsb_i\cdot\bsr+p\;\cos(\frac{4\pi}{5}i)+\frac{\pi}{3})$, with
	$\pg = 5\cdot\frac{\pi}{3} = \frac{-\pi}{3}$. The potential pattern remains self-similar. This degree of freedom does not couple to $\pg$ and to translations in the physical space.
	f) One of the 14-fold symmetric pattern that can be found for $\pg=0$ in the 7-fold symmetric configuration with $N=7$ (system size $3.8\lambda\times3.8\lambda$). Here $|\boldsymbol{b}_i|=1.95\;\frac{2\pi}{\lambda}$, as obtainable with seven beams by interference of every beam pair with $\pm\frac{6\pi}{7}$ relative angle.}
    \label{fig:quasicrystal}
\end{figure*}

Tunable trimers were previously realized \cite{Barter2020}, but as a 2D optical lattice and without passive stability. The example discussed here demonstrates that a 3D multi-frequency lattice can realize optical lattices with a basis of more than two sites, combining tunability and passive stability as in the 2D case demonstrated in the main text. 

\section*{Appendix B: Geometry phase in quasi-periodic potentials}

We demonstrate that the geometry phase is a concept relevant also for the description of quasi-periodic potentials, and show that its effect on the 2D quasiperiodic potential geometry resembles strikingly the situation in the hexagonal lattice. Furthermore, we show that the geometry phase can be tuned also in quasiperiodic-potentials via a multi-frequency design.

We consider the five-fold rotationally symmetric potential of the form:
\begin{equation}
     \Vpot(\bsr)=\sum_i^5\cos(\bsb_i\cdot\bsr+\phi_i)\label{eq:potential}
 \end{equation} 
where the reciprocal lattice vectors $\bsb_i$ and $\bsb_{i+1}$ have a relative angle of $\frac{2\pi}{5}$ [Fig.~\ref{fig:quasicrystal}(b)].

This potential can be seen as an incommensurate projection from a 4D periodic potential, because only four of the $\boldsymbol{b}_i$ are incommensurate. This means that the five $\phi_i$ control two degrees of freedom associated to translations in the physical dimensions, two degrees of freedom associated to translations in the not physically accessible dimensions (known as phasons), and finally, also a geometry phase of the form: 
\begin{equation}
    \phi_g=\sum_i^5 \phi_i
\end{equation}
We derive in Appendix C the general expression for $\pg$ as a function of the $\phi_i$ for any dimension and any number of 1D lattices making up the potential. 

We find that the quasi-periodic geometry indeed depends on the geometry phase $\pg$, and many qualitative features resemble its effect in the honeycomb lattice: for $\pg\sim2m\pi$, with $m$ integer, many shallow minima can be found and for $\pg\sim(2m+1)\pi$ fewer deeper minima can be found (Fig.~\ref{fig:quasicrystal}). The narrower distribution of on-site energies in the case $\pg=0$ can still be related to the frustration in momentum space, where a flux of $\Phi=\pi+\pg$ arises in a pentagonal plaquette [Fig.~\ref{fig:quasicrystal}(b)]. 

Also, in the region around the sign change of $\pg$, an emergence of an offset between even and odd local minima in approximately ten-fold symmetric structures can be observed, reminiscent of the breaking of the inversion symmetry in the boron-nitride lattice. A similar behaviour is observed also for the 7-fold symmetric quasicrystal lattice [Fig.~\ref{fig:quasicrystal}(e)].

The potential of Eq.~(\ref{eq:potential}) could be realized with the multi-frequency lattice, by taking 5 beams in the $xy$ plane with wave vectors $\boldsymbol{k}_i$ rotated by $i \times 2\pi/5$. Each beam is modulated such that it interferes only with the other two beams incoming with a relative angle $\pm4\pi/5$. This multi-frequency scheme would allow to tune dynamically $\pg$, as it provides control over each $\phi_i$.

In principle with this choice of the lattice beams, letting each beam pair give rise to interference \cite{Corcovilos2019} one gets $N=10$ wavevectors and hence a 6 dimensional $\pg$ (Appendix C). In that scheme, the 6 dimensional geometry phase is determined by the beam polarization, i.e., stable against phase fluctuations of the laser beams, but difficult to tune. In contrast, dynamical tuning of this complex object could be realized with a multi-frequency scheme (making use of ten different frequencies in total), while retaining the intrinsic stability of the geometry. 

The first realization of ultracold atoms in a quasicrystal optical lattice was done in an 8-fold symmetric optical lattice built from four 1D lattice rotated by $i \times \pi/4$ \cite{Viebahn2019}. From the considerations in Appendix C, one finds that the dimension of the geometry phase is zero in this case. The four phases of the 1D lattices only affect the position of the lattice and the two phasonic degrees of freedom. The quasicrystal lattice with multi-frequency realization proposed here is therefore the first scheme for a quasicrystal lattice with a tunable geometry.

While the phasonic degrees of freedom can be controlled via modulation of the laser beam phases \cite{Corcovilos2019}, as e.g. used in charge pump protocols \cite{Kraus2012,Bandres2016,Corcovilos2019,Marra2020} or in phasonic spectroscopy \cite{Rajagopal2019}, a multi-frequency realization of quasicrystal lattices further increases the modulation possibilities with e.g. quenches of the lattice geometry \cite{Flaschner2016}, or allowing for selective preparation of excited states via sweeps of $\pg$, as demonstrated in Fig.~\ref{fig:bandstructure}. 
Furthermore, theory work for ultracold bosons in quasicrystal optical lattices found a rich phase diagram strongly dependent on the local variation in the number of nearest neighbors \cite{Johnstone2019}, suggesting a pronounced dependence on the geometry phase in the case of the 5-fold symmetric quasicrystal lattice proposed here. Finally, the geometry phase controls the width of the distribution of on-site energies, which is an important tuning parameter for studies of the Bose glass or many-body localization \cite{Khemani2017}. 

These insights evidence how the geometry phase is a powerful concept for capturing important features of potentials like level statistics and breaking of local symmetries which even applies to non-periodic potentials.

\section*{Appendix C: Derivation of the geometry phase}

The derivation of Eq.~(\ref{eq:3beam}) is straightforward. The three beams in complex field notation shall be given by
\begin{align}
    \boldsymbol{E}_i(\boldsymbol{r},t) = \boldsymbol{E}_i^{(0)} e^{i (\boldsymbol{k r} - \omega t)} \mathrm{,} \label{eq:E(r,t)}
\end{align}
with $i = 1,2,3$ and the complex amplitudes $\boldsymbol{E}_i^{(0)}$ which contain field amplitude, polarization, and phase. The potential landscape generated by the three beams is proportional to the intensity of their total electric field and the intensity in turn is proportional to the absolute value squared of the complex electrical field resulting in
\begin{align}
    V(\boldsymbol{r},t) &= \kappa |\sum_{i=1}^3 \boldsymbol{E}_i(\boldsymbol{r},t)|^2 \\
    &= \kappa \sum_{i=1}^3 |\boldsymbol{E}_i(\boldsymbol{r},t)|^2 + \kappa \sum_{i=1}^3 2 \Re \left[ \boldsymbol{E}_i(\boldsymbol{r},t) \boldsymbol{E}_{i+1}^*(\boldsymbol{r},t) \right] \mathrm{,}
\end{align}
with proportionality constant $\kappa$ and $\boldsymbol{E}_4(\bsr,t) = \boldsymbol{E}_1(\bsr,t)$. The first term is independent of position and time and can be identified with $V_0$ in Eq.~(\ref{eq:3beam}). Plugging in Eq.~(\ref{eq:E(r,t)}) leads to
\begin{align}
    V(\bsr,t) &= V_0 + \kappa \sum_{i=1}^3 2 \Re \left[ \boldsymbol{E}_i^{(0)}{\boldsymbol{E}}_{i+1}^{*(0)} e^{i (\bsk_i - \bsk_{i+1})\bsr} \right] \\
    &= V_0 + \kappa \sum_{i=1}^3 2 \Re \left[ I_i^{(0)}e^{i\phi_i} e^{i (\bsk_i - \bsk_{i+1})\bsr} \right] \\
    &= V_0 + \kappa \sum_{i=1}^3 2 I_i^{(0)} \cos\left[(\bsk_i - \bsk_{i+1})\bsr + \phi_i \right] \mathrm{,} \label{eq:almost3beam}
\end{align}
where we introduced $I_i^{(0)} e^{i\phi_i} = \boldsymbol{E}_i^{(0)}\boldsymbol{E}_{i+1}^{*(0)}$. When defining $V_i = \kappa I_i^{(0)}$, Eq.~(\ref{eq:almost3beam}) becomes identical to Eq.~(\ref{eq:3beam}).

It should be stressed that the phases of the 1D lattices $\phi_i$ describe the geometry of the resulting lattices independently of how they arise from the interference of the lattice beams, which depends on the polarisation and detuning of the lattice. As an example, recall that a honeycomb lattice can be realized by a red-detuned lattice with in-plane polarization or a blue-detuned lattice with out-of-plane polarization \cite{Becker2010} and both situations are described by the same $\phi_i$.

We can identify the geometry phase as the single parameter describing the geometry of the lattice in the following way: consider a displacement of the lattice given by $\bsr \rightarrow \bsr + \delta\bsr$. Using $\sum_{i=1}^3 \bsb_i = 0$ we obtain that
\begin{align}
    \pg \rightarrow \pg + \delta\bsr \cdot\sum_{i=1}^3 \bsb_i = \pg
    \label{eq:shift}
\end{align}
is unchanged. From that we can conclude that the 2D parameter space corresponding to lattice displacements is orthogonal to the 1D parameter space spanned by $\pg$ which must therefore control the geometry.

This argument can be extended to any number of 1D lattices $N$ and physical dimensions $d$. Note that for quasiperiodic potentials, $D>d$, where $D$ is the dimension of the periodic potential whose incommensurate projection on the physical dimensions generates the quasiperiodic potential. $D$ is given by $D=N-N_\mathrm{iis}$, where $N_\mathrm{iis}$ is the number of rationally independent integer sequences $n_i^{(c)},...,n_N^{(c)}$, with $c=1...N_\mathrm{iis}$, such that
\begin{equation}
\sum_i^N\;n_i^{(c)}\boldsymbol{b}_i=0 \mathrm{.}
\label{eq:components}
\end{equation}
Because there are $N$ degrees of freedom associated to the $\phi_i$, and $D$ degrees of freedom associated to generic translations, the number of components of $\phi_g$ is given by $N_\mathrm{iis}=N-D$. The expression for the $c^\mathrm{th}$ component of $\pg$, $\phi_{g,c}$ can be directly obtained as
\begin{equation}
    \phi_{g,c}=\sum_i^N n_i^{(c)}\phi_i \mathrm{.}
\end{equation}

This expression can be verified as before by noting that $\phi_{g,c}$ is invariant under a generic translation $\bsr' \rightarrow \bsr' + \delta\bsr'$ in the $D-$dimensional space: 
\begin{align}
    \phi_{g,c} \rightarrow \phi_{g,c} + \delta\boldsymbol{r}' \cdot\sum_{i=1}^N \;n_i^{(c)}\bsb'_i = \phi_{g,c} \mathrm{,}
\end{align}
and hence must be a geometric object; $\boldsymbol{b}'_i$ are the $D-$dimensional wavevectors in the extended space.

Note that if $\sum_{i=1}^N \;n_i^{(c)}\boldsymbol{b}_i=0$, also $\sum_{i=1}^N \;n_i^{(c)}\boldsymbol{b}'_i=0$. In fact, only $D$ of the $\boldsymbol{b}'_i$ can be taken to be independent, and the remaining ones have to be defined as integer sums of the independent vectors such that Eq.~(\ref{eq:components}) is satisfied (for every sequence $c$). By this construction,  $\sum_{i=1}^N \;n_i^{(c)}\boldsymbol{b}'_i=0$ also in the not physically accessible dimensions. 

An alternative demonstration, which does not use the higher-dimensional space, can be made by noting that using the $\boldsymbol{b}_i$, weighted by the respective $n_i^{(c)}$, a plaquette in momentum space can be constructed with a flux $\Phi$ given by:
\begin{equation}
    \Phi=\sum_i n_i^{(c)}(\pi+\phi_i)=\sum_i n_i^{(c)}\pi+\phi_{g,c}
\end{equation}
and because the flux is gauge-invariant, $\phi_{g,c}$ must be related to the system geometry.

\section*{Appendix D: Spatial variation of the lattice geometry}

When the condition $\sum_i \bsb_i=0$ for the hexagonal lattice is not exactly met, the geometry phase becomes spatially dependent with a variation given by $\pg(\bsr)=\pg(0)+\sum_i \bsb_i \cdot \bsr$ [compare to Eq.~(\ref{eq:shift})].
In the multi-frequency lattice, because each 1D lattice is characterized by a different frequency, the corresponding wave vector is slightly different in magnitude from the others, causing $\sum_i \bsb_i\neq0$. In our realization with frequency differences in the MHz range, $|\sum_i \bsb_i|=0.3~$mrad/mm resulting in a negligible variation of $\pg$ over the system size. However, when working with modulation frequencies in the GHz range, one could engineer relevant spatial variations of the lattice geometry which could be exploited, e.g., for creating interfaces between system parts characterized by different topology \cite{Goldman2016}.

\section*{Appendix E: Implementation of the multi-frequency lattice}

\begin{figure}
	\includegraphics[width=0.8\linewidth]{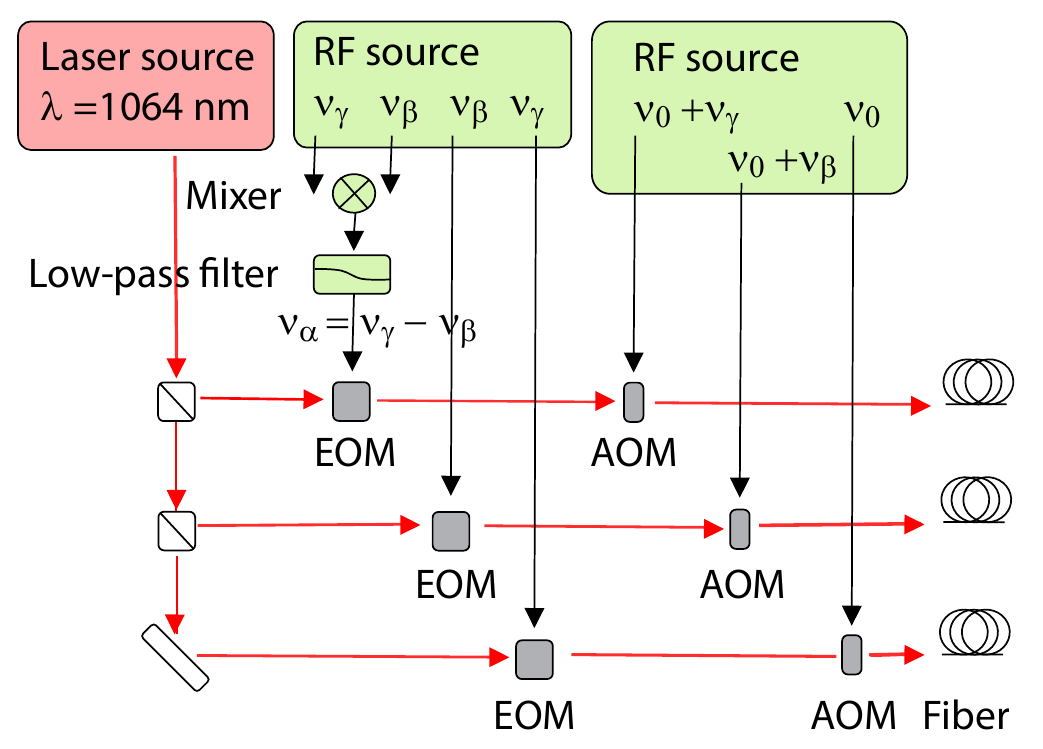}
	\caption{Implementation of the multi-frequency lattice with stable phases between the frequency components. The laser beam from a single source is split into three beams using $\lambda/2$ waveplates (not shown) and polarising beam splitters. Each beam has an EOM to add a sideband and an AOM to shift the frequency appropriately before coupling into separate optical fibers. The RFs for the EOMs are derived from a single source and one frequency is derived as the difference frequency of the other two using a mixer and a low pass filter. This setup produces the spectra sketched in Fig.~\ref{fig:scheme}(b).}
    \label{fig:implementation}
\end{figure}

Our lattice setup consists of a single laser source at $\lambda=1064$\,nm whose light is split into three beams. Each of them has an electro-optical modulator (EOM) (resonant high-Q electro-optic phase modulator from Qubig), adding sidebands to the laser spectrum at its respective driving frequency ($\nu_{\alpha}=2.22~$MHz, $\nu_{\beta}=7.77~$MHz, $\nu_{\gamma}=9.99~$MHz). The individual beams are shifted via the +1st order of acousto-optical modulators (AOMs) driven at frequencies $\nu_0+\nu_{\gamma}$, $\nu_0+\nu_{\beta}$ and $\nu_0=105.005~$MHz, (Fig.~\ref{fig:implementation}), ensuring that every pair of beams has exactly one frequency in common [Fig.~\ref{fig:scheme}(b)].

The modulation frequencies $\nu_{\alpha}$, $\nu_{\beta}$, $\nu_{\gamma}$ where chosen as multiple of 1.11~MHz in order to assure that the smallest frequency difference of all combinations is 1.11~MHz, i.e., higher than the typical energy scales of the atoms in the lattice. Such interference patterns form rapidly moving optical lattices that wash out and allow a description as non-interfering beams. Furthermore, this choice allows restricting the range of AOM frequencies to a band of 10~MHz. All interference terms from higher-order sidebands of the EOMs that give rise to static lattices are suppressed by at least a factor of $10^5$. The fast-moving part of the potential needs however to be taken into account, when calculating the external confinement. 

The stability of the geometry phase depends crucially on the stability of the RF sources used for the EOMs.
This can be seen by writing explicitely the time-dependence of the 1D lattice phases, as obtained after modulation of the laser beams with AOMs and EOMs (compare to Fig.~\ref{fig:scheme}), and without further assumptions: \begin{equation}\begin{aligned}
    \partial_t\phi_1=\;&(\nu_c-\nu_\alpha)-\nu_b\\
    \partial_t\phi_2=\;&(\nu_b-\nu_\beta)-\nu_a\\
    \partial_t\phi_3=\;&(\nu_a+\nu_\gamma)-\nu_c\\
\end{aligned}
\label{phi_i_timederivative}
\end{equation}

It follows that the time variation of $\phi_g$ is given by \begin{equation}
    \partial_t\phi_g=\sum_i^3\partial_t \phi_i=\nu_\gamma-\nu_\alpha-\nu_\beta
\end{equation}
Hence $\phi_g$ is constant as long as:
\begin{equation}\nu_\alpha=\nu_\gamma-\nu_\beta \label{lock_condition} \end{equation}
In order to ensure this condition, we derive the RF signals for the EOMs from the same digital frequency source. Furthermore, we derive the smallest frequency $\nu_{\alpha}$ as difference frequency $\nu_{\alpha}=\nu_{\gamma}-\nu_{\beta}$ from mixing the outputs of two channels at frequencies $\nu_\beta$ and $\nu_\gamma$ and using a low pass filter (Fig.~\ref{fig:implementation}). 
Thus perfect phase stability between the two channels operating at $\nu_\beta$ and between the two channels at $\nu_\gamma$ ensure that the condition of Eq.~(\ref{lock_condition}) is satisfied. 

The laser beams are overlapped in a plane under 120$^{\circ}$ to each other at the position of the atoms after passing through separate optical fibres (Fig.~\ref{fig:implementation}). Since the carrier frequency of each beam is the same as a 1st sideband frequency of another beam (e.g. beam $i$ and $i'$), the depth of the resulting 1D lattices is proportional to $J_0(n_i)J_1(n_{i'})$, where $J_0,\,J_1$ are Bessel functions of the first kind and $n_i,\,n_{i'}$ modulation indices of the corresponding beams. In order to maximize the depth of all three 1D lattices simultaneously, the modulation index at the EOMs thus has to be $n_{\alpha}=n_{\beta}=n_{\gamma}\approx1.08$.

We work with out-of-plane laser polarization, for maximal interference. This also avoids any vector light shifts, i.e., $m_F$ dependent potentials, which usually appear for a red-detuned honeycomb lattice \cite{Soltan-Panahi2011}. This choice also avoids possible Raman resonances to other $m_F$ states, which are not desired in this work, as opposed to optical Raman lattices for realizing spin-orbit coupling \cite{Wu2016,Wang2021Science}. We also made sure that the driving frequencies are not resonant with such transitions. 

The implementation of the multi-frequency lattice only requires additional EOMs and a change of the AOM frequencies on top of conventional setups of hexagonal optical lattices. It can therefore be a feasible way to achieve stable and tunable lattices in many experimental setups. 

\section*{Appendix F: Calibration of the lattice geometry}

While we can control the phases of the RF $\phi_\mathrm{rf}$ at the corresponding sources, the phase at the EOMs is shifted due to delays in the RF setup. We present different options in order to calibrate the connection between RF phase at the source $\phi_\mathrm{rf}$ and the lattice geometry, i.e., to find the value of $\phi_\mathrm{rf}$, denoted $\phi_\mathrm{rf}^0$, for which one gets a honeycomb lattice ($\pg=0$). In these measurements we vary the phase $\phi_\mathrm{rf}$ of a single RF signal ($\nu_{\alpha}$ in Fig.~\ref{fig:implementation}), while keeping the others constant.

\begin{figure}[!]
	\includegraphics[width=0.9\linewidth]{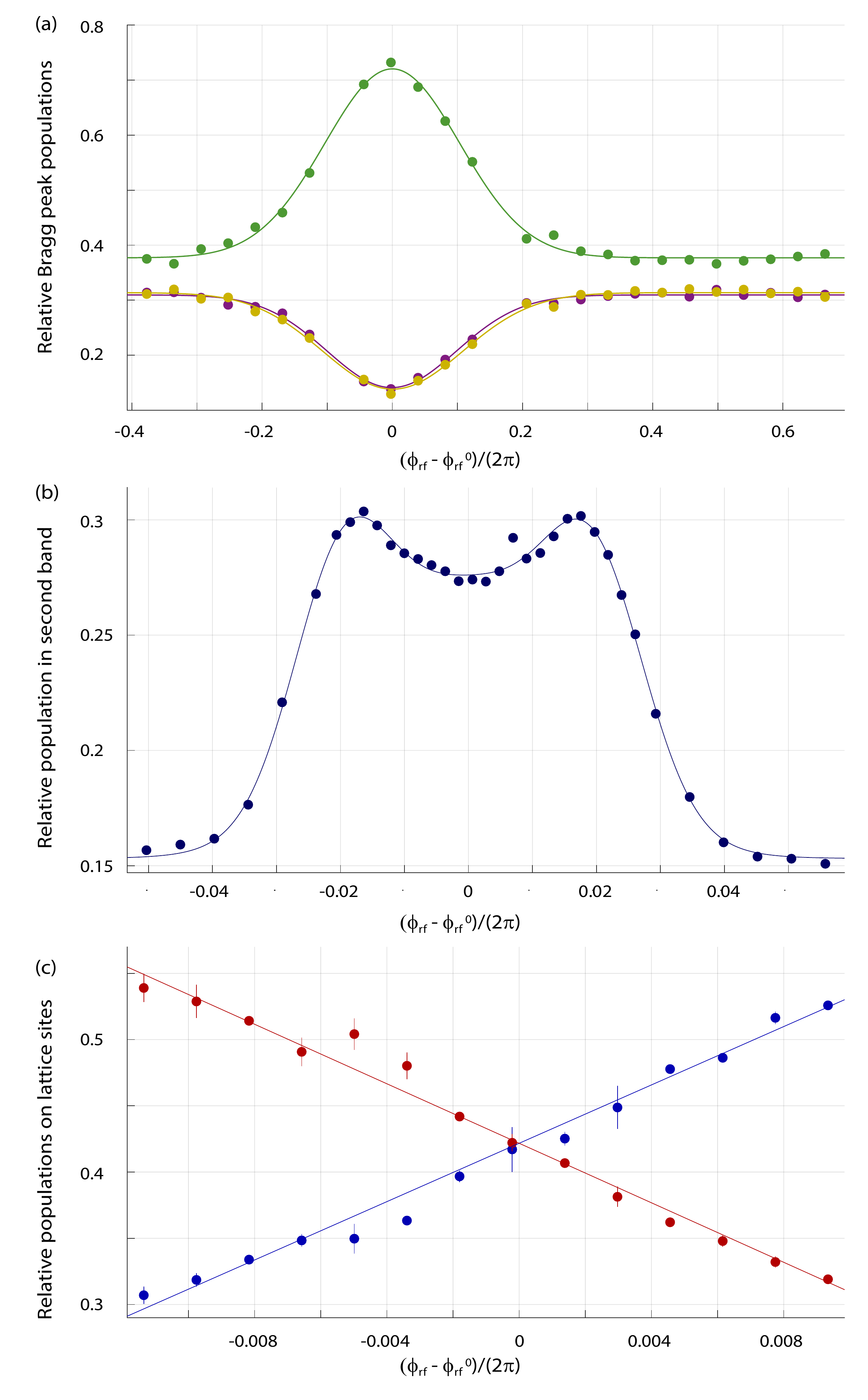}
	\caption{Different methods for the calibration of the geometry phase. (a) Relative strengths of imbalanced Bragg-peaks across all geometries. Around $\pg=0$ (honeycomb) the Bragg-peak populations are much more sensitive to the lattice imbalance (see text). The center is determined by the average of three heuristic Gaussian fits to the data, resulting in a 68\% confidence interval of the fit of $\delta\phi_\mathrm{rf}^0=0.00153\times2\pi=9.58~$mrad. The symbols correspond to experimental data, the lines to the fits. The color encodes the direction in momentum space. (b) Spectroscopy measurement closely around $\pg=0$ followed by band mapping, which yields the relative population of the 2nd band (symbols). A spectroscopy frequency resonant to a small AB-offset is used, resulting in two distinctive peaks at opposite sites of $\pg=0$. The center is determined by a heuristic three Gaussian fit with forced equal distances to each other (line). The 68\% confidence interval of the fit is then $\delta\phi_\mathrm{rf}^0=0.00016\times2\pi=0.98~$mrad. (c) Real-space measurement closely around $\pg=0$, with relative A-site (blue symbols) and B-site populations (red symbols). The error bars denote the standard deviation from 2 to 3 iterations. The populations do not add up to one due to finite occupations attributed to the spaces between the lattice sites. Via a heuristic fit of two linear curves (lines, valid only up to small offsets) the point of equal population $\pg=0$ is determined with a 68\% confidence interval of the fit of $\delta\phi_\mathrm{rf}^0=0.00012\times2\pi=0.75~$mrad.}
    \label{fig:calibration}
\end{figure}

A coarse calibration can be done by observing the asymmetry of Bragg peaks after time-of-flight of a BEC from a lattice with unbalanced lattice beam intensities. The asymmetry has a broad maximum around the honeycomb configuration ($\pg=0$)  [Fig.~\ref{fig:calibration}(a)]. In contrast to the momentum-space quantum walk of Fig.~\ref{fig:quantum-walk}, we adiabatically load into the lattice here and imbalance the lattice via the intensities of the lattice beams. Nevertheless, the larger response of the observed asymmetry for $\pg=0$ can be understood by the staggered magnetic flux $\Phi=\pi+\pg$ introduced above. The unbalancing of the lattice beam intensities would tend to create larger occupations of the momentum modes at multiples of one reciprocal lattice vector $\bf{b}_1$ in the ground state. This is, however, in competition with delocalization over the different momenta, i.e., similar populations of momentum modes at multiples of all three reciprocal lattice vectors, which lowers the energy. With the staggered flux around $\Phi=\pi$ in the honeycomb lattice at $\pg=0$, the system becomes frustrated and the delocalization is less energetically favorable, leading to the observed asymmetry in the momentum distribution. 

A second calibration method is band spectroscopy around $\pg=0$. By performing, e.g., sublattice modulation for a fixed frequency one gets a double-peaked signal for the relative population of the 2nd band versus $\pg$. The peaks are in correspondence to the resonance with the sublattice offset $\Delta_\mathrm{AB}$, and $\phi_\mathrm{rf}^0$ is found in the middle of the resonances [Fig.~\ref{fig:calibration}(b)].

A third calibration method is to measure the atomic populations on A- and B-sites across the honeycomb configuration $\pg=0$, with $\pg=0$ being at the intersection of the two populations.  This method is suitable for a precise calibration in a small region around $\pg=0$, where both sublattices have a significant population. This is a quick and precise calibration method, when single-site resolved imaging is available, e.g., via a matter-wave microscope \cite{Asteria2021} [Fig.~\ref{fig:calibration}(c)], and it is used for the stability analysis below. 

After determination of $\phi_\mathrm{rf}^0$, the calibration of the total lattice potential is completed by calibrating the depths of the 1D lattices $V_i$. This is done by standard techniques like, e.g., Kapitza Dirac scattering or spectroscopy. Throughout the manuscript, we state the lattice depth in units of the recoil energy $E_{\rm rec}=h^2/(2m\lambda^2)$, where $h$ is Planck's constant, $m$ is the mass of an $^{87}$Rb atom, and $\lambda=1064$~nm is the wavelength of the lattice beams.

\section*{Appendix G: Stability of the geometry phase}

We characterize the stability of $\phi_\mathrm{rf}^0$ by repeated calibration measurements of the geometry phase as in Fig.~\ref{fig:calibration}(c) every 26 minutes. They yield a standard deviation of $\delta\phi_\mathrm{rf}^0=3\,{\rm mrad}=0.17^{\circ}$, which we state as the relevant stability of the geometry phase. The individual measurements for the calibrations were taken every 36\,s and their standard deviation around the fitted curves is $\delta\phi_\mathrm{rf}^0=0.8\,{\rm mrad}=0.05^{\circ}$. These fluctuations might be limited by the 14 bit precision of our digital frequency source, i.e., a phase resolution of $0.00006 \times 2\pi=0.02^{\circ}$. 

\begin{figure}
	\includegraphics[width=0.8\linewidth]{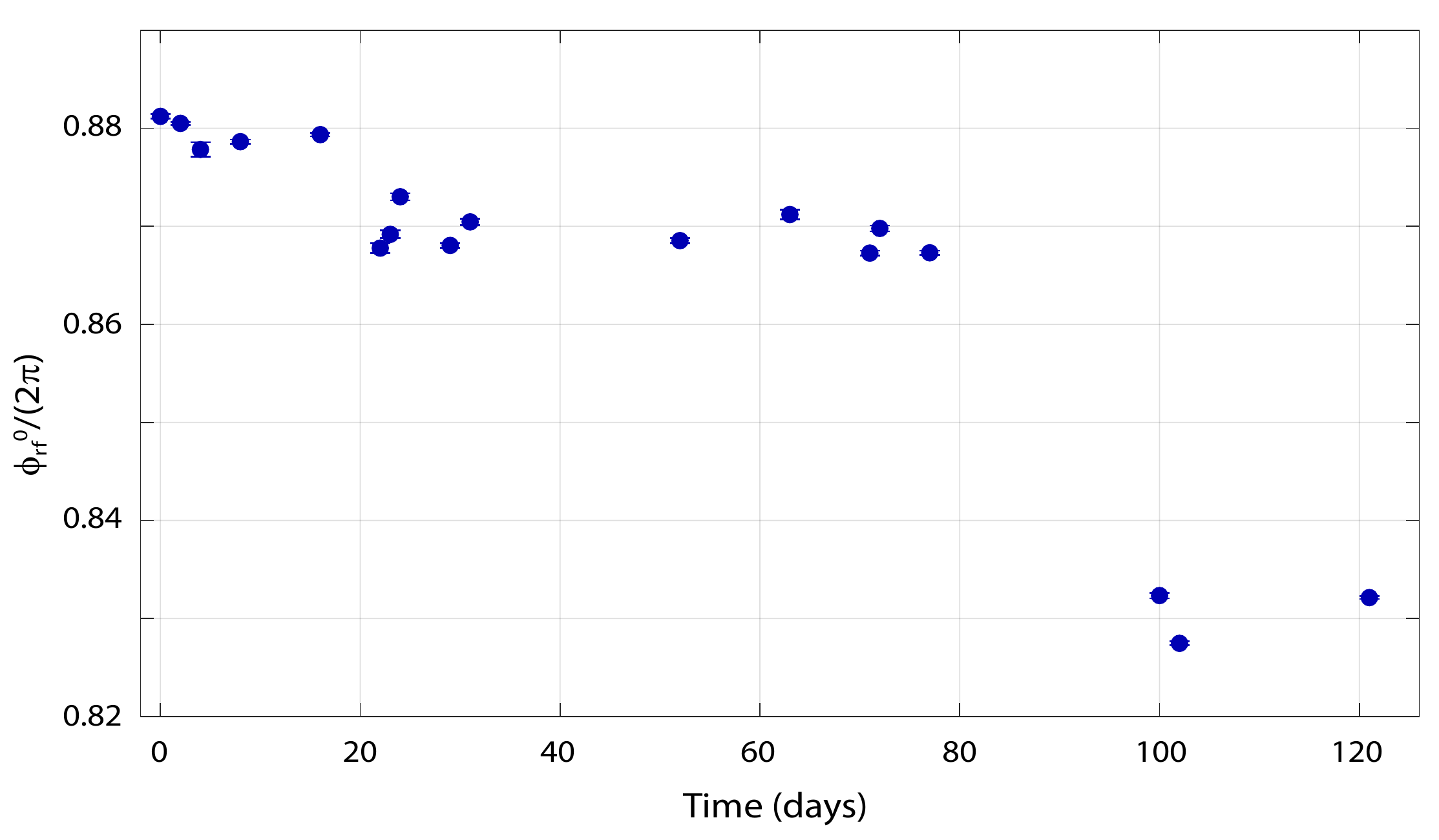}
	\caption{Evolution of the geometry phase calibration over four months. Every point results from a real-space phase calibration as in Fig.~\ref{fig:calibration}(c) with the error bar given by the 68\% confidence interval of the corresponding fit.}
    \label{fig:drift-geometry-phase}
\end{figure}

The long-term stability of the geometry-phase calibration over four months shows only very slow drifts with a stability within around $2^{\circ}$ for 60 days and only isolated jumps in the calibration (Fig.~\ref{fig:drift-geometry-phase}), probably due to changes in the laboratory conditions such as temperature or humidity. We find that the temperature stabilization of EOM crystals is important for ensuring this high passive stability both for the calibration of the lattice depth and for avoiding phase shifts from a mismatch of the temperature-dependent resonance condition of the resonant EOMs. For systems with phase locks, the long term stability usually requires regular recalibrations.

The geometry phase is given by the phases of the EOM sidebands alone and we therefore do not expect relevant phase noise on short time scales as it would arise on the phases of the laser beams from mechanical couplings. For reference, we give the phase noise of other setups of superlattices. Passively stable setups reach rms fluctuations of $25\,{\rm mrad}=1.4^{\circ}$ using different angles through a high-resolution objective \cite{Boll2016} or $1.5\,{\rm mrad}=0.1^{\circ}$ using a dual-wavelength interferometer \cite{Li2021}. For superlattice designs with active phase stabilization, phase drifts require a regular recalibration of the geometry. Additionally, the lock introduces phase noise, which can be reduced from $6.8^{\circ}$ \cite{Becker2010} and $5^{\circ}$ \cite{GreifPhD} down to $0.5^{\circ}$ \cite{LohsePhD} and $0.1^{\circ}$ \cite{Robens2017}. 

Next to the stability of the lattice geometry, the stability of the lattice position can also be of interest. Superlattice setups with phase locks typically control the geometry by keeping two lattices stable in absolute position. In the multi-frequency lattice, drifts of the lattice position are decoupled from drifts of the lattice geometry, the latter being only controlled by the phase of the EOM sidebands. In our setup, where the phases of the laser beams are not stabilized, we find position drifts between individual experimental shots \cite{Asteria2021}, while the lattice geometry remains very stable.

\section*{Appendix H: Implementation and modelling of the quantum walk}

For realizing short evolution times in the $\mus$ regime in a well-controlled lattice depth, we first ramp up the lattice beams and stabilize their intensity, while detuning the beams via the AOMs to avoid interference. We then start the pulsed evolution time in the lattice by setting the AOM frequencies to resonance and end it by setting the AOM amplitude to zero. The lattice could alternatively be switched via the EOMs, but we keep them always on to avoid thermal phase shifts that would affect the geometry phase. 

\begin{figure}
	\includegraphics[width=0.95\linewidth]{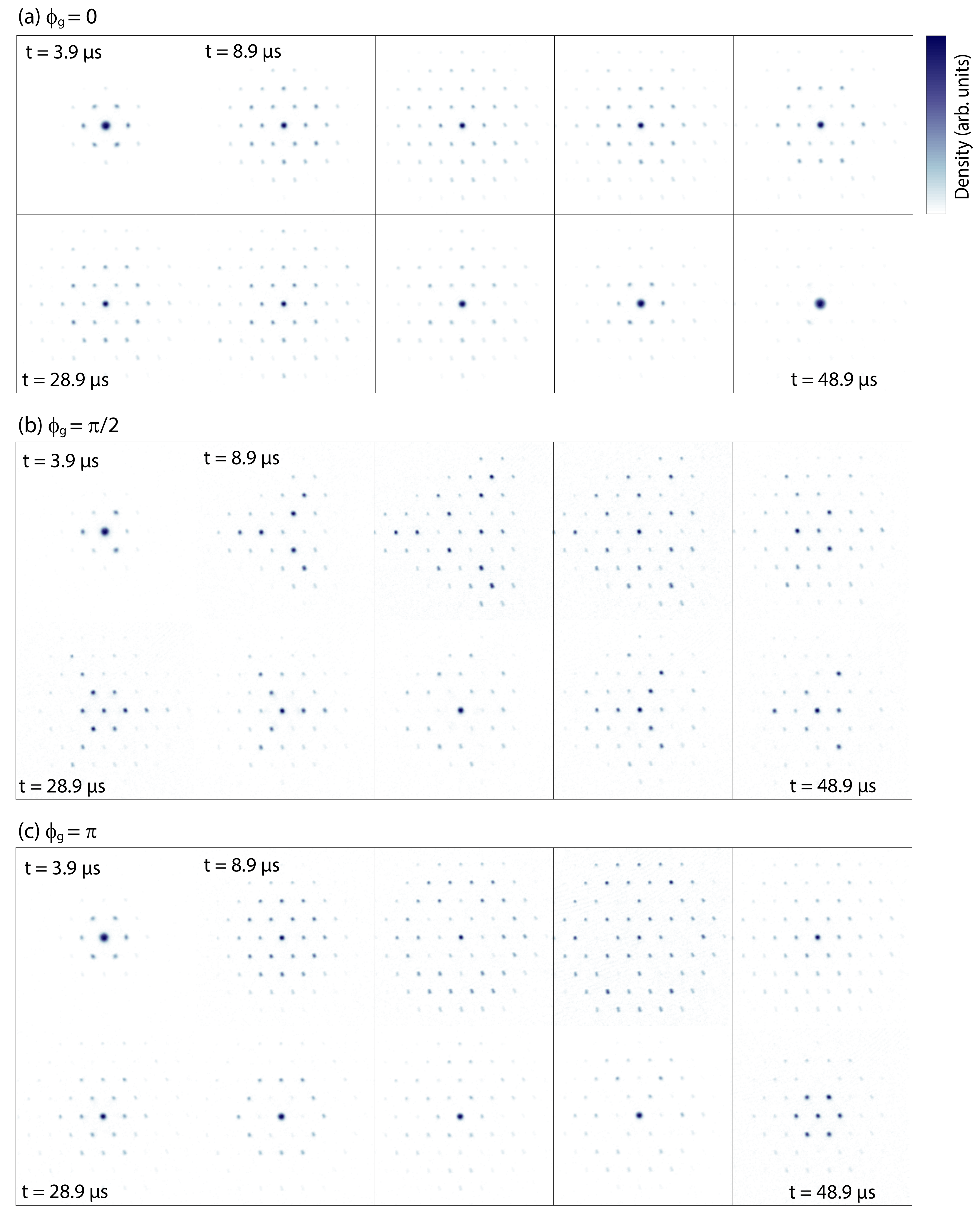}
	\caption{Quantum walk in a triangular momentum space lattice. (a) $\pg=0$, (b) $\pg=\pi/2$, (c) $\pg=\pi$. Each subfigure contains 10 different hold times, starting from 3.9\,µs and increasing in steps of 5\,µs up to 48.9\,µs. The lattice depth is the same as in Fig.~\ref{fig:quantum-walk}.}
    \label{fig:quantum-walk-data}
\end{figure}

Time-resolved data of the quantum walk from Fig.~\ref{fig:quantum-walk} is shown in Fig.~\ref{fig:quantum-walk-data}. The symmetry breaking observed in Fig.~\ref{fig:quantum-walk} can be captured by the simple calculation of the coherent dynamics in the lattice.

The numerical simulation shown in Fig.~\ref{fig:quantum-walk} is implemented as follows. We carry out exact diagonalization of the Hamiltonian matrix corresponding to the non-interacting lattice Hamiltonian in plane wave basis taking into account $11 \times 11$ reciprocal lattice sites. Using the resulting eigenstates $\ket{\psi_n}$ and energies $E_n$ we can compute the plane wave occupations as
\begin{align}
    n_{\boldsymbol{k}}(t) = \left| \sum_n \braket{\boldsymbol{k} | \psi_n} e^{-iE_nt/\hbar} \braket{\psi_n | \boldsymbol{k} = 0} \right|^2 \mathrm{,}
\end{align}
assuming the plane wave $\ket{\boldsymbol{k} = 0}$ as initial state.

\section*{Appendix I: Implementation and discussion of the sublattice modulation}

We modulate the three phases of the sideband frequencies $\nu_{\rm a}$, $\nu_{\rm b}$, $\nu_{\rm c}$ such that the modulation only couples to the lattice geometry, but not to a spatial translation of the lattice. This is achieved by symmetric modulation of the three 1D lattices 
\begin{equation}
    \phi_{\rm a,b,c}(t)=\phi_{\rm a,b,c}(0)\pm\frac{\delta \pg}{3}\sin(\omega t) \label{eq:modulation}
\end{equation} 
with a negative sign for $\phi_{\rm c}$ due to the opposite sideband used for the interference. 

The experiments start with a BEC of $^{87}$Rb atoms, which thermally fill up the lowest band after loading into the optical lattice. We modulate for 10 ms with modulation indices in the linear response regime. Finally we probe the system via band mapping (Appendix J) and integrate over the respective Brillouin zones to determine the band occupations.

\begin{figure}
	\includegraphics[width=0.95\linewidth]{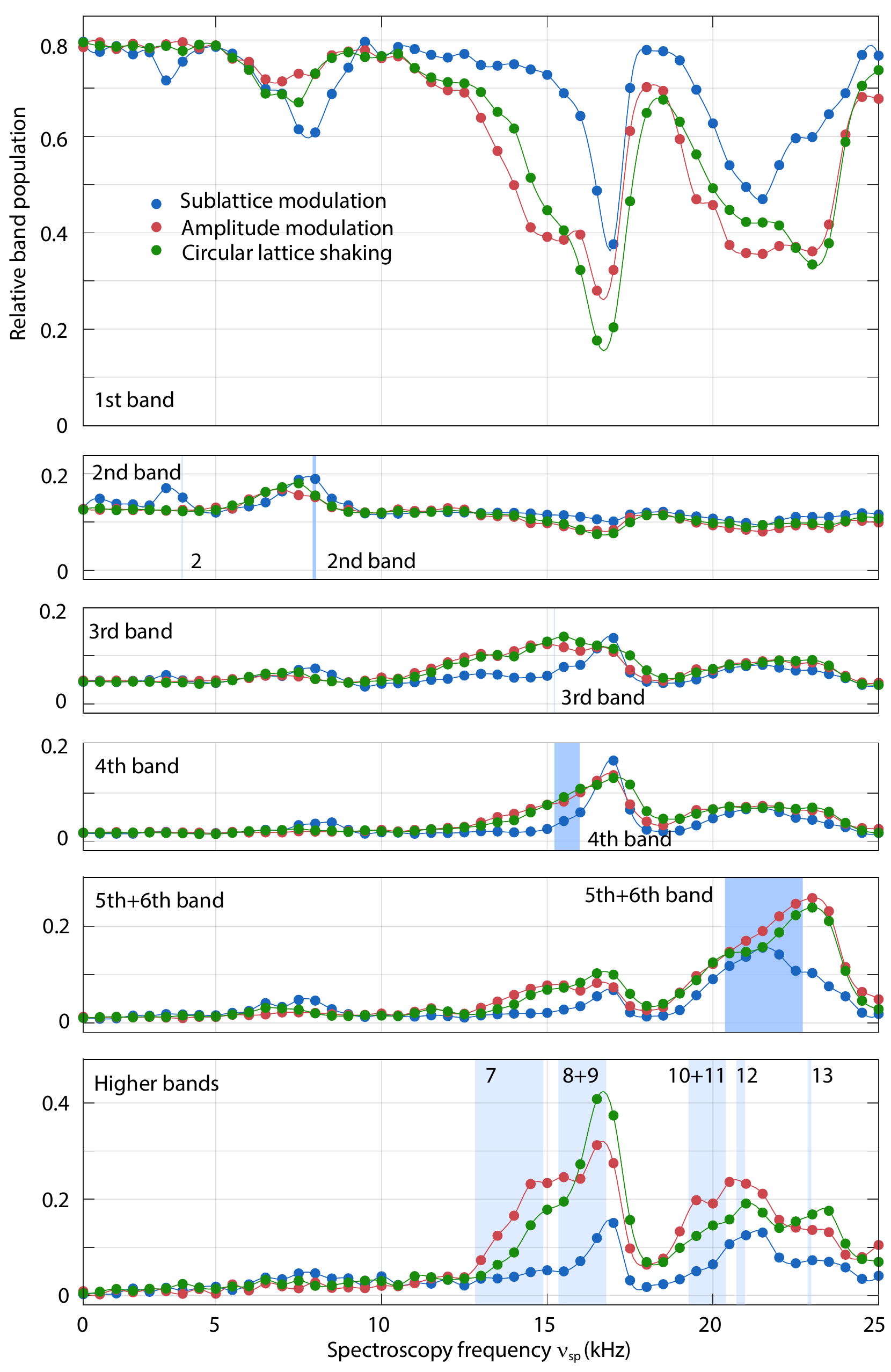}
	\caption{Band-resolved excitation spectra of a boron-nitride lattice with $\pg/(2\pi) = 0.024$. Relative populations of the different bands (see labels of the panels) versus spectroscopy frequency for sublattice modulation (blue), amplitude modulation (red) and circular lattice shaking (green). The shaded areas indicate the expected transition frequencies into the stated bands for single-photon transitions (blue) and two-photon transitions (light blue). The parameters for the data set are stated in Fig.~\ref{fig:spectroscopy}.}
    \label{fig:spectroscopy-full}
\end{figure}

For a detailed discussion of the spectra, we plot the occupations of all the higher bands, allowing to better identify the excitations (Fig.~\ref{fig:spectroscopy-full}). The excitation to the 2nd band around 7~kHz can be clearly identified in all three spectra. The sublattice modulation additionally has a two-photon transition to the 2nd band. The excitations to the connected 3rd and 4th bands around 16~kHz and to the connected 5th and 6th bands around 22~kHz are also present in all three spectra. Additionally, there are excitations around 15~kHz for the amplitude modulation and lattice shaking, which are absent for the sublattice modulation. We interpret them as two-photon transitions into the seventh band and assume that the observed smaller signal in the 3rd to 6th bands are due to band decay during the excitation pulse. Similarly, there are two-photon transitions to the connected 10th and 11th bands around 20~kHz and to the 13th band around 23~kHz, which explain the broader excitation feature for amplitude modulation and lattice shaking. We conclude that the spectra for sublattice modulation are cleaner and potentially better suited for selective coupling to higher bands or off-resonant Floquet protocols.

\begin{figure}
	\includegraphics[width=1\linewidth]{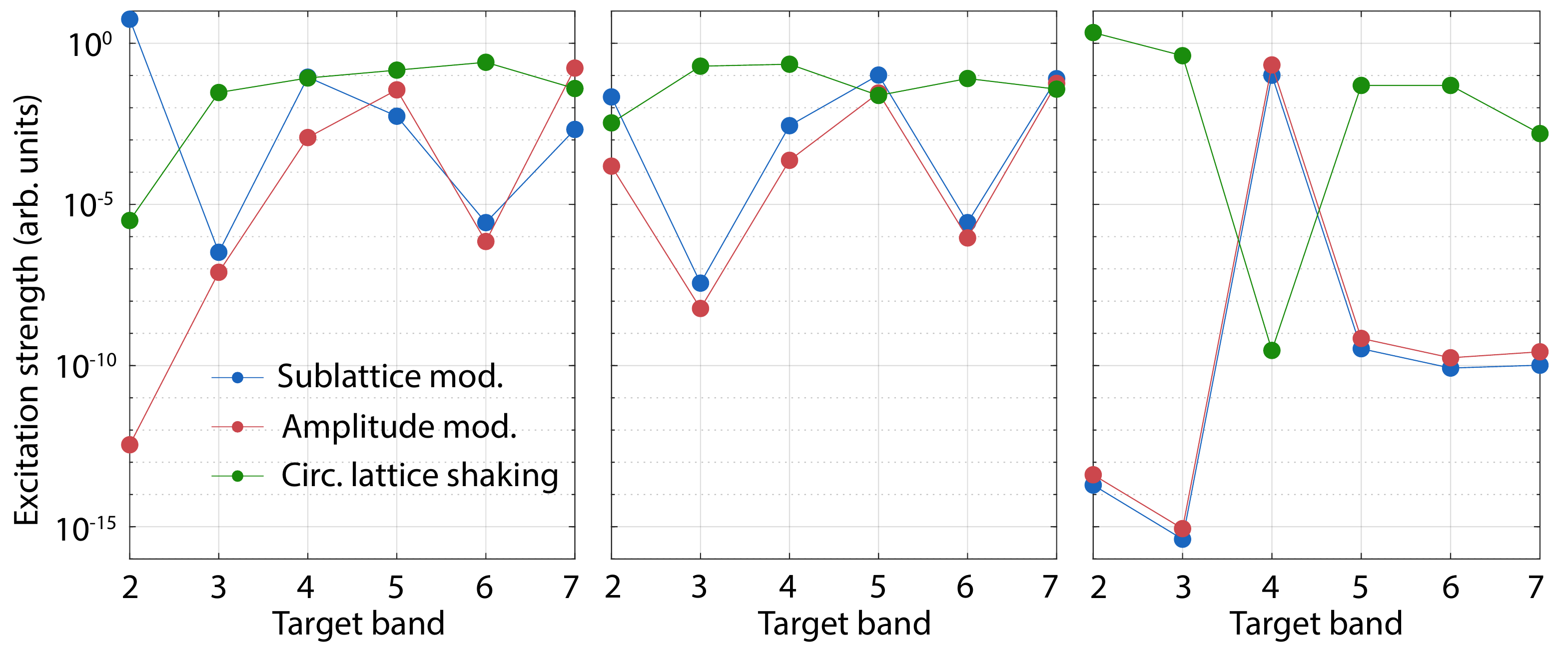}
	\caption{Resonant excitation strength from the lowest band. Transition matrix elements into higher bands for sublattice modulation (blue), amplitude modulation (red) and circular lattice shaking (green) integrated over the first Brillouin zone plotted on a logarithmic scale. The examined geometries are from left to right: $\pg/(2\pi) = 0, 0.024, 0.5$. The modulation strengths correspond to the experimental values of Fig.~\ref{fig:spectroscopy}, i.e., $\epsilon_\mathrm{sm}=0.02$ for sublattice modulation, $\epsilon_\mathrm{am}=0.05$ for amplitude modulation and $\epsilon_\mathrm{cs}=0.08$ (resonance to 2nd band), 0.035 (resonance to 3rd and 4th band), 0.028 (resonance to 5th and 6th band), 0.021 (theoretical expectation for resonance to 7th band) for circular lattice shaking.}
    \label{fig:excitation-strengths}
\end{figure}

In order to interpret these observations, we evaluate the matrix elements of the different perturbations, as relevant for linear response theory, and integrate them over quasimomentum (Appendix K). We plot the resulting excitation strengths integrated over the first Brillouin zone versus band number for different geometry phases and find clear differences between the three modulation types (Fig.~\ref{fig:excitation-strengths}). From the simple analogy to a harmonic oscillator, which couples all levels for shaking but only levels with the same parity for amplitude modulation, we expect a stronger band dependence for amplitude modulation and sublattice modulation. Indeed, these modulations have strongly suppressed excitation strengths to the 3rd and 6th band. The excitation strengths to the 2nd band, however, differ between amplitude modulation and sublattice modulation: in the honeycomb lattice only the sublattice modulation efficiently couples into the 2nd band.

The experimental observations roughly match with this simple consideration, while a complete quantitative understanding of the spectra would probably also need to include the complete dynamics during the excitation pulse as well as interaction effects, which can be sizable also in a lattice of tubes \cite{Ozawa2017}. In particular, the larger ratio between the excitation strengths to the 2nd and the 4th band for sublattice modulation compared to amplitude modulation for $\pg=0.15$ as in the measurements, fits with the experimental observation. The suppressed excitation strengths into the 3rd and 6th band (Fig.~\ref{fig:excitation-strengths}) are not easily resolved in the experiment, because these bands are connected to the 4th and 5th band, respectively, and one expects significant excitation into the combined bands.
Furthermore, the two-photon resonance to the 2nd band for the sublattice modulation is in agreement with the higher-order terms that arise in the expansion of the perturbation for the sublattice modulation (also compare the analysis of higher-order terms for phasonic modulation \cite{Rajagopal2019}). 

\section*{Appendix J: PCA analysis of distorted Brillouin zones}

\begin{figure}[!]
	\includegraphics[width=0.95\linewidth]{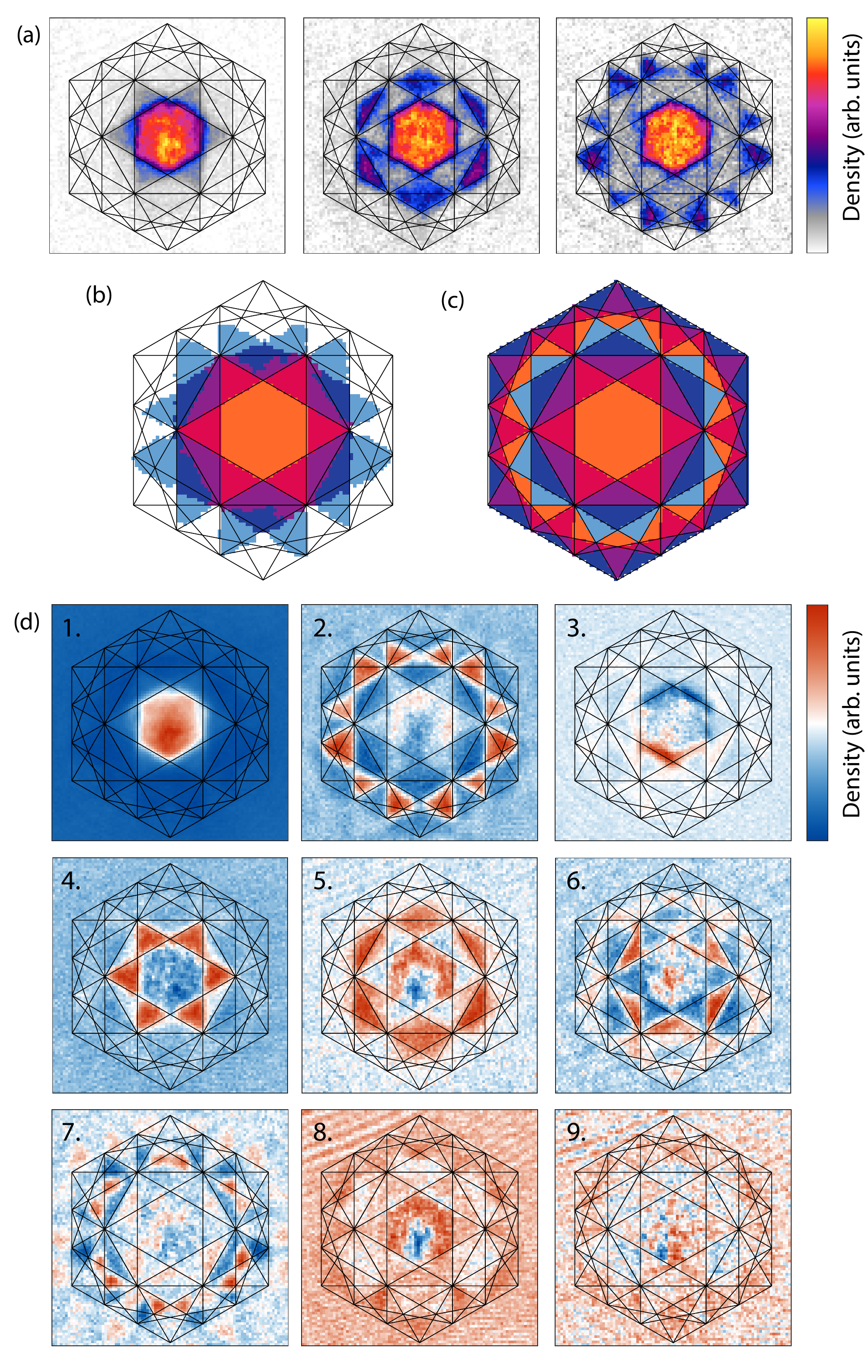}
	\caption{PCA analysis of distorted higher BZs. (a) Example band mapping images of atoms in higher bands indicate distortions relative to the ideal BZs (black lines). (b) Masks of the first five distorted BZs obtained by the PCA analysis of the band mapping images. (c) The first nine ideal BZs sampled with the same pixel size for comparison. (d) The first nine PCs of the data set of 147 images as in (a) carry information on the distorted BZs.}
    \label{fig:pca-analysis}
\end{figure}

We analyze the band populations via adiabatic band mapping \cite{Kohl2005}, i.e., by exponentially ramping down the lattice depth within $1.5~\rm ms$ such that the higher bands are mapped onto the higher Brillouin zones (BZ) of the hexagonal lattice. The optical confinement from the lattice beams with an estimated trapping frequency of 65~Hz is ramped down along with the lattice depth and the additional magnetic confinement with trapping frequency of 45~Hz is switched off at the end of the lattice ramp within 40~µs. In the resulting images, we find a reproducible distortion of the populated regions compared to the ideal BZs [example images in Fig.~\ref{fig:pca-analysis}(a)]. These distortions are particularly pronounced for higher BZs, which become rounded and shifted in the direction of gravity, i.e., towards the bottom in the figure. We expect this to result from the effect of the external trap, possible misalignment of the magnetic and optical confinement, and gravity during band mapping. Note that in contrast to other setups, the direction of gravity lies within the 2D lattice plane in our system.

Because these effects are difficult to simulate numerically, we extract the distorted BZs from the data itself by performing a principal component analysis (PCA) on the combined set of measurements from Fig.~\ref{fig:spectroscopy} (147 images in total). The BZ masks resulting from the analysis explained below [Fig.~\ref{fig:pca-analysis}(b)] are then used for the extraction of the relative populations in the spectroscopy.

PCA has been previously used in cold atom experiments to extract collective modes, noise patterns, or density fluctuations of Bose-Einstein condensates in a model-free way \cite{Dubessy2014, Cao2019, Hertkorn2021} as well as for fringe removal algorithms \cite{Song2020, Xiong2020}. The mathematical protocol can be found e.g. in ref.~\cite{Dubessy2014}. The eight most relevant principal components (PCs) are shown in Fig.~\ref{fig:pca-analysis}(c). Higher orders only contain noise.

These PCs can be used to identify the distorted BZ and to derive the corresponding masks. We keep the ideal BZs for the 1st and 2nd band, because their deformations are very small. To construct the 3rd to 5th+6th BZ masks, we pick the PC with the most contrast between the corresponding expected region and its surroundings. We first determine the 5th+6th BZ mask using the 2nd PC. By binarizing with the threshold at the middle of the range of values (white color in the image) while forcing pixels inside the 1st and 2nd BZs to be zero, we get a distorted mask, which fits well to the images with 5th+6th band excitations (see below). 

From this we also get the combined region of the distorted 3rd and 4th BZs as everything between the 2nd and 5th+6th BZ masks. At last, we determine the border between the 3rd and 4th BZs using the 4th and 6th PC. We use the sum of both, after normalizing them to span the same range of values, as we find this to have less fuzzy boundaries, but we get very similar results for separate analyses of the PCs. After this analysis, there are still some islets of a few pixels sprinkled in the oppositely assigned regions, which have sizes larger 50 pixels, which we flip setting an appropriate threshold. While these BZ masks are derived from the data set of the spectroscopy, they also fit reasonably well to the images obtained by sweeping into higher bands of Fig.~\ref{fig:bandstructure} considering the differences in lattice depth. 

The 5th+6th BZ mask obtained in this way is almost twice as large as the other four masks. We conclude that it contains the 5th and 6th BZ. This is supported by the observation that these two bands touch [compare Fig.~\ref{fig:bandstructure}(b)] and stay connected during the lattice ramp down of the band mapping protocol. It was found in previous work on checkerboard lattices that the populations of BZs mix even when the bands only touch briefly during the lattice ramp \cite{Wirth2011, Hachmann2021}. In our system, this happens e.g. between the 2nd and 4th band, which might explain the residual population of the 2nd BZ in the third image of Fig.~\ref{fig:bandstructure}(a). The shape of the distorted 5th+6th BZ mask does indeed contain much of the ideal 6th BZ, even though other parts of it are missing. It contains almost nothing of the 7th BZ, which fits to the gap between the 6th and 7th band down to small lattice depths. For a complete understanding of the band mapping images, a full numerical calculation of the lattice ramp including the harmonic trap would be worthwhile although beyond the scope of this article.

\section*{Appendix K: Calculation off matrix elements}

We derive the resonant excitation strengths for the three different types of lattice modulations presented in Fig.~\ref{fig:spectroscopy}. In the case of amplitude modulation every lattice beam intensity $I_i$ is modulated periodically with equal frequency $\omega/(2\pi)$, phase and strength $\epsilon_{\rm am}$:
\begin{equation*}
I_i(t)=I_{i,0}+\epsilon_{\rm am} I_{i,0} \sin{(\omega t)}.
\end{equation*}
Restricting also to the balanced case ($I_{1,0}=I_{2,0}=I_{3,0}$) this leads to the total lattice potential
\begin{align}
	V_{\rm am}(\bsr,\pg,t)&=2V(1 + \epsilon_{\rm am}\sin(\omega t)) \sum_{i} \cos\Bigl(\bsb_i \cdot \bsr + \frac{\pg}{3} \Bigr) \notag \\
	&= \Vpot(\bsr,\pg) + \sin(\omega t) \Vpot'(\bsr,\pg).
\end{align}
with the time-independent perturbation operator $\Vpot'(\bsr,\pg)=\epsilon_{\rm am} \Vpot(\bsr,\pg)$.  

We calculate the excitation strength of this time-dependent potential using time-dependent perturbation theory. According to Fermi's golden rule, the excitation rate $\Gamma^{\bsq,\bsq'}_{B,B'}$ from an initial quasimomentum $\bsq$ in $B=1$ to a final quasimomentum $\bsq'$ in band $B'$ is given by 
\begin{equation}\label{eq:ProbFMGR}
\Gamma^{\bsq\bsq'}_{B,B'}\propto\bigg|\int{d^3r\bsp^{\bsq}_B(\bsr)\Vpot'(\bsr)\bsp^{\bsq'*}_{B'}(\bsr)}\bigg|^2.
\end{equation}
The states $\bsp$ are eigenstates of the static lattice Hamiltonian $\hat{H}$, given by periodic Bloch states
\begin{equation}\label{eq:BlochStates}
\bsp_{B}^\bsq(\bsr)=e^{-i\bsq\cdot\bsr}\sum_{u}c_{u}^{\bsq,B}e^{-i\bsk_u\cdot\bsr},
\end{equation}
with Bloch coefficients $c_{u}^{\bsq,B}$ and $\bsk_u$ being integer linear combinations of the reciprocal lattice vectors $\bsb_i$. Plugging one summand of $\Vpot'(\bsr)$ e.g. $V'_1=\epsilon_{\rm am} 2 V \cos(\bsb_1 \bsr + \pg/3)$ and (\ref{eq:BlochStates}) into equation (\ref{eq:ProbFMGR}) we get

\begin{widetext}
\begin{align} 
\int{d^3r\bsp^{\bsq}_B V'_1\bsp^{\bsq'*}_{B'}} &= \epsilon_{\rm am}V
\int{d^3 re^{-i(\bsq-\bsq')\cdot\bsr} 
\left(\sum_{u}{c^{\bsq,B}_u e^{-i\bsk_u\cdot\bsr+i\bsb_1\cdot\bsr+i\pg/3}}\right)  
\left(\sum_{u'}{c^{*\bsq',B'}_{u'} e^{i\bsk_{u'}\cdot\bsr}}\right)} \notag \\
&+ \epsilon_{\rm am}V 
\int{d^3r e^{-i(\bsq-\bsq') \cdot\bsr} 
\left(\sum_{u}{c^{\bsq,B}_u e^{-i\bsk_u\cdot\bsr-i\bsb_1\cdot\bsr-i\pg/3}}\right)
\left(\sum_{u'}{c^{*\bsq',B'}_{u'} e^{i\bsk_{u'}\cdot\bsr}}\right)} \notag \\
&= \epsilon_{\rm am}V \delta_{\bsq,\bsq'} \sum_{u,u'} \bigg(e^{i\pg/3} \delta_{\bsk_u+\bsb_1,\bsk_{u'}} + e^{-i\pg/3} \delta_{\bsk_u-\bsb_1,\bsk_{u'}}  \bigg) c^{\bsq,B}_u c^{*\bsq',B'}_{u'} \notag \\
&= \epsilon_{\rm am}V \delta_{\bsq,\bsq'} \left(\sum_{u}{c^{*\bsq,B'}_u \bigg( c^{\bsq,B}_{\bsk_u+\bsb_1} e^{-i\pg/3} + c^{\bsq,B}_{\bsk_u-\bsb_1} e^{i\pg/3} \bigg)} \right). 
\end{align}
Doing the analogous for all summands of $\Vpot'(\bsr)$ we get for the rate
\begin{align}
\Gamma^{\bsq\bsq'}_{B,B'} &\propto \delta_{\bsq,\bsq'}  \epsilon_{\rm am}^2V^2 \Bigg| 
   \left(\sum_{u}{c^{*\bsq,B'}_u \bigg( c^{\bsq,B}_{u+\bsb_1} e^{-i\pg/3} + c^{\bsq,B}_{u-\bsb_1} e^{i\pg/3} \bigg)} \right) \notag \\
&+ \left(\sum_{u}{c^{*\bsq,B'}_u \bigg( c^{\bsq,B}_{u+\bsb_2} e^{-i\pg/3} + c^{\bsq,B}_{u-\bsb_2} e^{i\pg/3} \bigg)} \right) +
   \left(\sum_{u}{c^{*\bsq,B'}_u \bigg( c^{\bsq,B}_{u+\bsb_3} e^{-i\pg/3} + c^{\bsq,B}_{u-\bsb_3} e^{i\pg/3} \bigg)} \right)
\Bigg|^2 \mathrm{.}
\end{align}

In the case of sublattice modulation, the phase of every 1D lattice is equally periodically modulated with amplitude $A_\mathrm{sm}$, ensuring a fixed lattice position and a change in $\pg$ of three times the phase amplitude [Eq.~(\ref{eq:modulation})]. This changes the perturbation operator with respect to the static lattice Hamiltonian.

\begin{align}
V_{\rm sm}(r,\pg,t)&=2V \left(\sum_{i} \cos\Bigl(\bsb_i \cdot \bsr + \frac{\pg}{3} + \frac{A_{\rm sm}}{3} \sin(\omega t)\Bigr)\right)  \notag \\
&= 2V \left(\sum_{i} \cos\Bigl(\bsb_i\cdot\bsr + \frac{\pg}{3}\Bigr) \cos\Bigl(\frac{A_{\rm sm}}{3} \sin(\omega t)\Bigr) - \sin\Bigl(\bsb_i\cdot\bsr + \frac{\pg}{3}\Bigr) \sin\Bigl(\frac{A_{\rm sm}}{3} \sin(\omega t)\Bigr) \right) \notag \\
&\approx 2V \left(\sum_{i} \cos\Bigl(\bsb_i\cdot\bsr + \frac{\pg}{3}\Bigr) + \cos\Bigl(\bsb_i\cdot\bsr + \frac{\pg}{3} + \frac{\pi}{2}\Bigr) \frac{A_{\rm sm}}{3} \sin(\omega t) \right) \notag \\
&= \Vpot(\bsr,\pg) + \Vpot(\bsr,\pg - \pi/2) \epsilon_{\rm sm} \sin(\omega t) = \Vpot(\bsr,\pg) + \sin(\omega t) \Vpot'(\bsr,\pg - \pi/2)
\end{align}

With $\epsilon_{\rm sm}=\frac{A_{\rm sm}}{3}$. In the third line we only used the first order of the sine and cosine with $\sin{\omega t}$ in the argument, as the perturbation is very small. Higher orders lead to the appearance of terms modulated with multiples of $\omega$. For the excitation probability this results only in an additional phase and a different prefactor compared to amplitude modulation. Thus the excitation rate for sublattice modulation is

\begin{align}
\Gamma^{\bsq\bsq'}_{B,B'} &\propto \delta_{\bsq,\bsq'}  \epsilon_{\rm sm}^2V^2 \Bigg| 
   \left(\sum_{u}{c^{*\bsq,B'}_u \bigg( c^{\bsq,B}_{u+\bsb_1} e^{-i(\pg/3+\pi/2)} + c^{\bsq,B}_{u-\bsb_1} e^{i(\pg/3+\pi/2)} \bigg)} \right)  \\
&+ \left(\sum_{u}{c^{*\bsq,B'}_u \bigg( c^{\bsq,B}_{u+\bsb_2} e^{-i(\pg/3+\pi/2)} + c^{\bsq,B}_{u-\bsb_2} e^{i(\pg/3+\pi/2)} \bigg)} \right) +
   \left(\sum_{u}{c^{*\bsq,B'}_u \bigg( c^{\bsq,B}_{u+\bsb_3} e^{-i(\pg/3+\pi/2)} + c^{\bsq,B}_{u-\bsb_3} e^{i(\pg/3+\pi/2)} \bigg)} \right) \notag
\Bigg|^2.
\end{align}

Lastly the excitation probability for the circular shaking is determined using our lattice description. Here two of the three lattice beams are periodically modulated with amplitude $\Delta\nu$ as $\delta\nu_1=0, \delta\nu_{2,3}=2\Delta\nu \big( \pm\cos(\omega t) + \sqrt{3}\sin(\omega t) \big)$, which is equivalent to detunings of the sideband frequencies $\nu_{\alpha, \beta, \gamma}$ of
\begin{align}
\delta\nu_{\alpha}&=\delta\nu_1-\delta\nu_2=-2\Delta\nu \big(\cos(\omega t) + \sqrt{3} \sin(\omega t)\big) \mathrm{,} \notag \\
\delta\nu_{\beta}&=\delta\nu_2-\delta\nu_3=4\Delta\nu \cos(\omega t) \mathrm{,} \\
\delta\nu_{\gamma}&=\delta\nu_3-\delta\nu_1=-2\Delta\nu \big(\cos(\omega t) - \sqrt{3} \sin(\omega t)\big) \mathrm{.} \notag
\end{align}
As this gives $\sum_{i} \delta\nu_i = 0$ it indeed only changes the lattice position and not its geometry. From the detuning we get the time derivative of the phase $\dot{\phi}=2\pi\delta\nu_i$ and thus
\begin{align}
\phi_{\alpha}&=A_{\rm cs} \big(\sin(\omega t) - \sqrt{3} \cos(\omega t)\big) \mathrm{,} \notag \\
\phi_{\beta}&=2A_{\rm cs} \sin(\omega t) \mathrm{,}  \\
\phi_{\gamma}&=A_{\rm cs} \big(\sin(\omega t) + \sqrt{3} \cos(\omega t)\big) \mathrm{,} \notag
\end{align}
with $A_{\rm cs}=4\pi\Delta\nu/\omega$. The time-dependent potential for circular shaking then is
\begin{align}
V_{\rm cs}(\bsr,\pg,t) =2V \bigg( &\cos\Bigl( \bsb_1\cdot\bsr + \frac{\pg}{3} - A_{\rm cs} \big( \sin(\omega t) - \sqrt{3} \cos(\omega t) \big) \Bigr) \notag \\ 
+ &\cos\Bigl( \bsb_2\cdot\bsr + \frac{\pg}{3} + 2A_{\rm cs}  \sin(\omega t) \Bigr) \notag \\
+ &\cos\Bigl( \bsb_3\cdot\bsr + \frac{\pg}{3} - A_{\rm cs} \big( \sin(\omega t) + \sqrt{3} \cos(\omega t) \big) \Bigr)  \bigg) \notag \\
\approx 2V \bigg( &\cos\Bigl( \bsb_1\cdot\bsr + \frac{\pg}{3} \Bigr) + \cos\Bigl( \bsb_1\cdot\bsr + \frac{\pg}{3} - \frac{\pi}{2} \Bigr) A_{\rm cs} \big( \sin(\omega t) - \sqrt{3} \cos(\omega t) \big) \notag\\
+ &\cos\Bigl( \bsb_2\cdot\bsr + \frac{\pg}{3} \Bigr) + \cos\Bigl( \bsb_2\cdot\bsr + \frac{\pg}{3} + \frac{\pi}{2} \Bigr) 2A_{\rm cs}  \sin(\omega t)  \notag\\
+ &\cos\Bigl( \bsb_3\cdot\bsr + \frac{\pg}{3} \Bigr) + \cos\Bigl( \bsb_3\cdot\bsr + \frac{\pg}{3} - \frac{\pi}{2} \Bigr) A_{\rm cs}  \big( \sin(\omega t) + \sqrt{3} \cos(\omega t) \big)   \bigg) \notag \\
= V_{\rm cs}(\bsr,\pg) + \epsilon_{\rm cs} 2V &\bigg( 
\frac{1}{2}\cos \Bigl( \beta_1-\frac{\pi}{2} \Bigr)  \big( \sin(\omega t) - \sqrt{3} \cos(\omega t) \big) 
+ \cos \Bigl( \beta_2+\frac{\pi}{2} \Bigr) \sin(\omega t)
+ \frac{1}{2} \cos \Bigl( \beta_3-\frac{\pi}{2} \Bigr)  \big( \sin(\omega t) + \sqrt{3} \cos(\omega t) \big) \bigg).
\end{align}
With $\epsilon_{\rm cs}=2A_{\rm cs}$. In the end we used $\beta_i=\bsb_i\cdot\bsr + \frac{\pg}{3}$ and again sine and cosine with $\sin(\omega t)$ in the argument are approximated to first order. Plugging the modulated part of $V_{\rm cs}(\bsr,\pg,t)$ into equation (\ref{eq:ProbFMGR}) we get
\begin{align}
&\Gamma^{\bsq\bsq'}_{B,B'} \propto \delta_{\bsq,\bsq'} \epsilon_{\rm sm}^2V^2 
\Bigg| (\frac{1}{2} + \frac{i\sqrt{3}}{2}) \left(\sum_{u}{c^{*\bsq,B'}_u \bigg( c^{\bsq,B}_{u+\bsb_1} e^{-i(\pg/3-\pi/2)} + c^{\bsq,B}_{u-\bsb_1} e^{i(\pg/3-\pi/2)} \bigg)} \right) \\
&\!\!\!\!+ \left(\sum_{u}{c^{*\bsq,B'}_u \bigg( c^{\bsq,B}_{u+\bsb_2} e^{-i(\pg/3+\pi/2)} + c^{\bsq,B}_{u-\bsb_2} e^{i(\pg/3+\pi/2)} \bigg)} \right)
 + (\frac{1}{2} - \frac{i\sqrt{3}}{2}) \left(\sum_{u}{c^{*\bsq,B'}_u \bigg( c^{\bsq,B}_{u+\bsb_3} e^{-i(\pg/3-\pi/2)} + c^{\bsq,B}_{u-\bsb_3} e^{i(\pg/3-\pi/2)} \bigg)} \right)
\Bigg|^2. \notag
\end{align}
\end{widetext}


\bibliographystyle{apsrev4-2}

\end{document}